\documentclass[usenatbib,a4paper,fleqn,twocolumn]{mnras}            

\usepackage{txfonts}
\usepackage{amssymb}	
\usepackage{graphicx} 
\usepackage{txfonts}
\usepackage{adjustbox}
\usepackage{caption}
\usepackage{float}
\usepackage[flushleft]{threeparttable}
\usepackage{longtable}
\usepackage{textcomp}
\usepackage{multirow}
\usepackage{hhline}
\usepackage{lscape}
\usepackage{pdflscape}
\usepackage{colortbl}
\usepackage{multicol}
\usepackage{natbib}

\title[VOICE shear bias calibration]{Weak Lensing Study in VOICE Survey II: Shear Bias Calibrations}

\author[Liu et al.]
{Dezi Liu$^{1,2,3}$\thanks{E-mail: adzliu@pku.edu.cn},
Liping Fu$^{2}$\thanks{E-mail: fuliping@shnu.edu.cn},
Xiangkun Liu$^{3}$,
Mario Radovich$^{4}$,
Chao Wang$^{1}$,
\newauthor
Chuzhong Pan$^{1}$,
Zuhui Fan$^{1}$\thanks{E-mail: fanzuhui@pku.edu.cn},
Giovanni Covone$^{5,6,7}$, 
Mattia Vaccari$^{8,9}$,
\newauthor
Maria Teresa Botticella$^{7}$,
Massimo Capaccioli$^{5}$,
Demetra De Cicco$^{5}$,
\newauthor
Aniello Grado$^{7}$,
Lance Miller$^{10}$,
Nicola Napolitano$^{7}$,
Maurizio Paolillo$^{5,6}$,
\newauthor
Giuliano Pignata$^{11,12}$
\\
\\
$^{1}$Department of Astronomy, School of Physics, Peking University, Beijing 100871, China\\
$^{2}$The Shanghai Key Lab for Astrophysics, Shanghai Normal University, 100 Guilin Road, 
          Shanghai 200234, China\\
$^{3}$South-Western Institute for Astronomy Research, Yunnan University, Kunming 650500, China \\
$^{4}$INAF--Osservatorio Astronomico di Padova, vicolo dell'Osservatorio 5,  Padova 35122, Italy\\
$^{5}$Dipartimento di Fisica ``E. Pancini", Universit\`a degli Studi Federico II, Napoli 80126, Italy \\
$^{6}$INFN, Sezione di Napoli, Napoli 80126, Italy\\
$^{7}$INAF--Osservatorio Astronomico di Capodimonte, Salita Moiariello 16, Napoli 80131, Italy\\
$^{8}$Department of Physics \& Astronomy, University of the Western Cape, Robert Sobukwe Road,
7535 Bellville, Cape Town, South Africa \\
$^{9}$INAF - Istituto di Radioastronomia, via Gobetti 101, 40129 Bologna, Italy\\
$^{10}$Department of Physics, Oxford University, Keble Road, Oxford OX1 3RH, UK\\
$^{11}$Departemento de Ciencias Fisicas, Universidad Andres Bello, Avda. Republica 252, Santiago, Chile\\
$^{12}$Millennium Institute of Astrophysics (MAS), Nuncio Monse\~nor S\'otero Sanz 100, Providencia, Santiago, Chile
}

\date{Accepted XXX. Received YYY; in original form ZZZ}

\pubyear{2018}

\begin{document}
\label{firstpage}
\pagerange{\pageref{firstpage}--\pageref{lastpage}}
\maketitle

\begin{abstract}
The VST Optical Imaging of the CDFS and ES1 Fields (VOICE) Survey is proposed 
to obtain deep optical $ugri$ imaging of the CDFS and ES1 fields using 
the VLT Survey Telescope (VST). At present, the observations for the CDFS field have 
been completed, and comprise in total about 4.9 deg$^2$ down to $r_\mathrm{AB}$\,$\sim$\,26\,mag.  
In the companion paper by \citet{2018arXiv180210282F}, we present the weak lensing shear 
measurements for $r$-band images with seeing $\le$ 0.9\,arcsec. In this paper, we perform 
image simulations to calibrate possible biases of the measured shear signals. 
Statistically, the properties of the simulated point spread function (PSF) and galaxies show good 
agreements with those of observations. The multiplicative bias is calibrated to reach an accuracy 
of $\sim$3.0\%. We study the bias sensitivities to the undetected faint galaxies 
and to the neighboring galaxies.  We find that undetected galaxies contribute to the 
multiplicative bias at the level of $\sim$0.3\%. Further analysis shows that galaxies with lower 
signal-to-noise ratio (SNR) are impacted more significantly because the undetected galaxies skew 
the background noise distribution. For the neighboring galaxies, we find that although most 
have been rejected in the shape measurement procedure, about one third of them still remain 
in the final shear sample. They show a larger ellipticity dispersion and contribute to $\sim$0.2\% of the 
multiplicative bias. Such a bias can be removed by further eliminating these neighboring galaxies. 
But the effective number density of the galaxies can be reduced considerably.  Therefore efficient 
methods should be developed for future weak lensing deep surveys.
\end{abstract}




\section{Introduction}
The inhomogeneous matter distribution in the Universe deflects gravitationally the light rays 
from distant galaxies, resulting in tiny shape and flux changes of their observed images. This 
phenomenon is usually referred to as the weak gravitational lensing, or cosmic shear (see e.g. 
\citet{2014RAA....14.1061F,2015RPPh...78h6901K,2017SchpJ..1232440B,2017arXiv171003235M} 
for recent reviews). The induced galaxy shape distortions reflect directly the gravitational tidal field, 
and hence contain valuable cosmological information. On the other hand, because of the weakness 
of the cosmic shear signals and the existence of the intrinsic ellipticities for galaxies,  the weak lensing 
studies are statistical in nature. We need to measure a large number of galaxies accurately.  The 
observational advances  have made the weak lensing effect a powerful cosmological probe 
\citep{2007MNRAS.381..702B,2013MNRAS.430.2200K,2013MNRAS.432.2433H,2016PhRvD..94b2001A,
2016PhRvL.117e1101L}. From the Canada-France-Hawaii Telescope Lensing Survey (CFHTLenS; 
\citet{2013MNRAS.432.2433H}) to the ongoing surveys, such as the Dark Energy Survey (DES; 
\citet{2016PhRvD..94b2002B,2016MNRAS.460.2245J,2017arXiv170801533Z}), the Hyper Suprime-Cam 
(HSC) Survey \citep{2012SPIE.8446E..0ZM,2018PASJ...70S..25M}, and the Kilo-Degree Survey 
(KiDS; \citet{2015MNRAS.454.3500K,2017MNRAS.465.1454H}), the survey area has increased from 
$\sim$200\,deg$^2$ to a few thousands square degrees. The future surveys, notably the ground-based 
Large Synoptic Survey Telescope (LSST;  \citet{2009arXiv0912.0201L}), and the space missions of 
Euclid \citep{2011arXiv1110.3193L} and the Wide Field Infrared Survey Telescope (WFIRST; 
\citet{2012arXiv1208.4012G}), will be able to further enhance the statistical power of weak lensing studies. 

Because the weak lensing induced shape distortion only accounts for a few per cent, much smaller 
than the intrinsic ellipticity of galaxies, observationally, weak lensing studies require accurate measurements. 
This is extremely challenging.  Several programs have devoted many endeavors to test the capabilities of 
different shear measurement algorithms, and to study their sensitivities to various systematics, such as the 
imperfect modeling of the variations of the PSF and the telescope observing conditions 
\citep{2006MNRAS.368.1323H,2007MNRAS.376...13M,2009AnApS...3....6B,2012MNRAS.423.3163K,
2014ApJS..212....5M}. 

In addition, the physical properties of galaxies themselves can also bias the shear measurements 
\citep{2017arXiv171003235M}. For example, the Gravitational Lensing Accuracy Testing 3 (GREAT3; 
\citet{2014ApJS..212....5M,2015MNRAS.450.2963M}) challenge investigated the impact of the complex 
galaxy morphology on the measured shear, and concluded that it can affect the calibration by about one 
per cent for many methods. \citet{2017MNRAS.468.3295H} also studied the sensitivity of shape measurements 
to other galaxy properties based on \textit{Euclid}-like image simulation, and highlighted the impact 
of galaxies below the survey detection limit. Another well known effect is the light contamination from 
neighboring galaxies. With the increase of the survey depth, such blending effect becomes increasingly 
a concern \citep{2018MNRAS.475.4524S,2017arXiv171000885M}.  As presented in \citet{2013MNRAS.429.2858M}, 
over 20\% of galaxies have neighbors in CFHTLenS, whose $i'$-band limiting magnitude is 
$i'_\mathrm{AB}=24.54$\,mag. These neighboring galaxies are generally excluded for shear 
measurements because the superposition of their isophotes can lead to large and biased 
ellipticity estimate. This exclusion does not significantly affect the cosmological studies using CFHTLenS due 
to their small fraction relative to the total galaxy sample. However, in the case of deeper observations, more 
galaxies are expected to suffer from blending effect \citep{2013MNRAS.434.2121C}. For example,
58\% of objects in the HSC Wide survey are blended \citep{2018PASJ...70S...5B}.
Simply excluding these blenders undoubtedly will reduce the effective number density of galaxies considerably and 
hence degrade the statistical power for cosmological studies. How to properly take into account the blending 
effect and make these galaxies usable in the shear analyses still remains to be a challenging task. 

In this paper, we perform image simulations based on the VOICE survey (PIs: Giovanni Covone 
\& Mattia Vaccari; \citet{2016heas.confE..26V}) for shear measurement calibrations. Together with 
the SUDARE survey \citep{2015A&A...584A..62C,2017A&A...598A..50B}, VOICE was proposed to 
cover about eight square degrees evenly split between the CDFS \citep{2001ApJ...551..624G,
2001ApJ...562...42T} and the ES1 \citep{2000MNRAS.316..749O,2004MNRAS.351.1290R} fields in four 
optical $ugri$ bands using VST/OmegaCam. The survey aims at providing deep optical images in the targeted 
fields to enable various astrophysical studies in conjunction with other existing data covering different 
wavelength \citep{2015fers.confE..27V,2016ASSP...42...71V}. One of the main scientific objectives is to detect 
galaxy clusters at intermediate redshifts and determine their two-dimensional mass distributions using the 
weak lensing shear signals of background galaxies. The imaging observations of CDFS field have been completed. 
Our shear measurements and image simulations then focus on this field. It was divided into four tiles (CDFS1--4), 
with each about one square degree. Over one hundred exposures, spanning almost two years, with a single 
exposure time of 360 seconds, were obtained for each tile \citep{2015A&A...579A.115F}. 
The observation was conducted in dithering mode made of five exposures per epoch to cover the detector gaps. 
For each epoch, the exposure times and dithering patterns were identical to those of the KiDS survey 
\citep{2015A&A...582A..62D}. The images were preprocessed (including flat fielding, cosmic ray removal etc.) with the 
VST-Tube pipeline \citep{2012MSAIS..19..362G}. Selecting only those images with a full width at half 
maximum (FWHM) less than 0.9 arcsec, the final mosaic reaches a $5\sigma$ limiting magnitude 
of $r_\mathrm{AB}$\,$\sim$\,26.1\,mag with $2\arcsec$ aperture diameter for point sources, 1.2\,mag deeper than 
KiDS. The galaxy shapes were measured using \texttt{LensFit} \citep{2007MNRAS.382..315M,2008MNRAS.390..149K,
2013MNRAS.429.2858M} on the $r$-band images (\citet{2018arXiv180210282F}; F18 hereafter). Our final shear catalog 
contains $\sim$$3.2\times10^5$ galaxies. The effective number density of galaxies is about 16.4\,arcmin$^{-2}$, a factor 
of two higher than that of the KiDS survey. 

The paper is organized as follows. In Section~\ref{sec:ShearMeasure}  we briefly introduce the shape 
measurements of galaxies in the VOICE survey. The image simulation procedures are detailed in 
Section~\ref{sec:ImgSim}. The bias calibrations of the measured shear are presented in 
Section~\ref{sec:BiasCal}.  The bias sensitivities, especially the impact of blending effect and undetected 
galaxies, are discussed in Section~\ref{sec:dis}.  We summarize our results in Section~\ref{sec:Con}. 
Note that all magnitudes in this paper are quoted in the AB system \citep{1983ApJ...266..713O}.

\section{Weak Lensing Shear Measurements}\label{sec:ShearMeasure}
In this section, we summarize the procedures of shear measurements for VOICE. More details can be 
found in F18. 

The single exposure images after astrometric and photometric calibrations are stacked for source detection 
and photometry using \texttt{SExtractor} package \citep{1996A&AS..117..393B}. The stars and galaxies are 
then separated by combining multi-band colors and the magnitude-size relation. In total, about 150,000 
galaxies and 2800 PSF stars are extracted in each tile. These PSF stars are selected to be brighter than 
22.0\,mag with SNR higher than 20 and have nearly uniform distribution over the entire images. 

The galaxy shapes are measured for each tile individually using \texttt{LensFit} which is a Bayesian model 
fitting code. The surface brightness distributions of galaxies are modeled as a de Vaucouleurs bulge plus an 
exponential disk components. In \texttt{LensFit}, the fitting for a galaxy is done on individual exposures.  
The ellipticity is then derived by combining the likelihoods of different exposures, with a marginalization over 
other free parameters (i.e. the galaxy position, scalelength, flux and bulge fraction) with the adopted priors 
\citep{2013MNRAS.429.2858M}. In the meantime, a weight is assigned to each galaxy which includes both 
the measurement uncertainty and the intrinsic ellipticity dispersion of galaxies. If a galaxy has an unsuccessful 
shape measurement, the corresponding weight is set to be zero. Each object is also flagged with an integer to 
indicate its characteristics, with a number of zero meaning a successful model fit to the galaxy. \texttt{LensFit} 
was originally optimized to measure the cosmic shear in CFHTLenS \citep{2012MNRAS.427..146H}, and at 
present has also been applied to other surveys, such as KiDS \citep{2015MNRAS.454.3500K,2017MNRAS.465.1454H} 
and RCSLenS \citep{2016MNRAS.463..635H}.

Accurate PSF modeling is crucial for weak lensing shear measurements.  In \texttt{LensFit}, 
the PSFs are determined as postage stamps of pixel values on each exposure individually based on the input 
star catalog. In this stage, \texttt{LensFit} firstly removes stars whose central pixel is more than half of the saturation 
level or have SNR smaller than 20. If any pixel in a star is flagged as ``bad" or belong to another object, it is also 
excluded. \texttt{LensFit} then computes a cross-correlation coefficient between the profile of a star and the local PSF model, 
obtained by a polynomial fitting. Only stars with the cross-correlation coefficient larger than 0.86 are used for the final PSF 
modeling. The distribution of the cross-correlation 
coefficient peaks at 1.0 with the median value of 0.97. In order to model the spatial variations of the PSF over the entire image 
mosaic, a forth-order polynomial fit is applied. In addition, a first-order chip-dependent polynomial is used to take 
into account the discontinuities in the PSF across the boundaries between CCDs. To further validate the PSF 
modeling, F18 calculated the star-galaxy cross-correlation function and found it generally consistent with zero.

To deblend the neighboring galaxies, \texttt{LensFit} creates isophotes after smoothing their surface brightness 
distributions with a Gaussian function of FWHM to be equal to that of the local PSF. If the isophotes of the target 
galaxy are touching with the neighbors at a given threshold (2$\sigma$ by default) above the smoothed pixel noise, 
these galaxies will be excluded from further analysis. Furthermore, if the centroid of a galaxy, measured from the 
pixels within the threshold in the stacked stamp, does not lie within 4 pixels around the target position in the original 
input galaxy catalog, it is also excluded. These galaxies either have close neighbors or are individuals with complex 
morphology. With the default threshold of 2$\sigma$, we find that about one third of galaxies have shape measurements 
with non-zero weight when only using the single epoch images. This is similar to KiDS results as expected. When adding 
data from more exposures, we expect that the number of successful shear measurements for galaxies should increase 
because of the increase of SNR for galaxies. However, this is not the case with the $2\sigma$ threshold.  This can be 
understood as follows. The default threshold in \texttt{LensFit} is optimized for CFHTLenS-like surveys, which are 
shallower than VOICE. In VOICE,  we have a larger number of faint detections, and the lower background noise compared 
to CFHTLenS makes the default $2\sigma$ contour larger, therefore leading to more rejections due to the presence of neighbors.
We have performed extensive tests, and found that changing the threshold to $5\sigma$ 
can lead the number of galaxies with non-zero weight to a reasonable level. Therefore, we use this $5\sigma$ threshold 
for shape measurement in both the observational analyses and simulation studies. 

Finally, over 300,000 galaxies in the entire field have shape measurements with ellipticity dispersion of about 0.298. 
Following the definition in \citet{2012MNRAS.427..146H}, the derived weighted number density is about 16.35 per 
arcmin$^2$ over the total effective sky coverage of 4.13\,deg$^2$ after rejecting the masked regions.

\section{Image Simulation}\label{sec:ImgSim}
We use \texttt{Galsim} \citep{2015A&C....10..121R}, a widely used galaxy image 
simulation toolkit, to create the simulated images. 
\texttt{Galsim} can generate star and galaxy images with 
specified analytic surface brightness profiles or based on direct \textit{HST} observations.  
Different image transformations and noise models can be efficiently handled by the software.
A framework for simulating weak lensing shear  is also encoded. 
In our studies, the simulation is performed in two steps for each tile. 
As a first step, we generate a mock catalog which contains the celestial coordinates, magnitudes, 
morphologies and ellipticity of the simulated objects. This mock catalog is then used to 
create single exposure images for shape measurements.

\subsection{Mock Catalogs}\label{sec:mcat}
In the simulation, we use the sources detected in the observed images as the parent 
sample, and fix their celestial coordinates and fluxes to the observed values. This 
takes into account in a natural way the galaxy clustering and blending effect. Following 
\citet{2013MNRAS.434.2121C}, we define the neighbors by their separation 
on the celestial sphere (further discussion on the definition is given in 
Section~\ref{sec:dis}). The fraction of neighboring galaxies within a given distance 
$r$ is shown in Table~\ref{tab:blend}. It is seen that the fraction increases 
significantly as the separation gets larger, reaching about 16\% for distances of $r\leq3.0\arcsec$. 
These neighbors can potentially bias the shape measurements. We note that galaxies fainter than the detection 
limit are missing in our parent sample, but  they may  also introduce biases in the measured 
cosmic shear \citep{2015MNRAS.449..685H,2017MNRAS.468.3295H,2017MNRAS.467.1627F}. 
For the VOICE survey, however, we find that their effects are almost negligible. Detailed 
investigation on these systematics will be presented in Section \ref{sec:dis}.
We do not include saturated stars in the parent sample. As shown in 
F18, they have been masked out before performing shape measurements.

As in \texttt{LensFit}, the galaxy profiles are modeled as a linear 
combination of a de Vaucouleurs bulge and an exponential disk. 
Following \citet{2013MNRAS.429.2858M}, the galaxy bulge to total flux ratio ($B/T$) is 
randomly sampled from a truncated Gaussian distribution $N(0.0, 0.1^2)$ in the range of 0.0 to 1.0, and 
around ten percent of galaxies are set to be bulge-dominated with $B/T=1.0$.
The intrinsic ellipticity as well as the size (defined as the disc scalelength along the major axis)  
distributions of the galaxies are kept to be the same as that in \citet{2013MNRAS.429.2858M} for 
CFHTLenS simulations. In the fiducial model, the dispersion of the intrinsic ellipticity is close to $\sigma_\mathrm{int}=0.25$. 
The relationship between the $r$-band magnitude and 
median disc scalelength involved in the size distribution follows the equation given by 
\citet{2015MNRAS.454.3500K}. These distributions also correspond to 
the \texttt{LensFit} priors used for VOICE shape measurements. The orientations of the galaxies are randomly 
assigned, following a uniform distribution on the interval [$-\pi/2$, $\pi/2$].

A constant shear with modulus $|g|=0.04$ is applied to all galaxies. To calibrate the measured shear 
signal to about one per cent level, in this case, the minimum number of simulated galaxies is required to be 
$n_\mathrm{gal}=[\sigma_\mathrm{int}/(0.01|g|)]^2\simeq3.9\times10^5$ \citep{2007MNRAS.376...13M}. 
As shown in the following section, our simulation can satisfy the criterion. As a compromise between deriving valid 
shear calibration and  saving computational time, four different shear 
combinations ($g_1$,\,$g_2$) are used, which are ($+0.0283$,\,$+0.0283$), ($-0.0283$,\,$-0.0283$), 
($-0.0370$,\,$+0.0153$), and ($+0.0153$,\,$-0.0370$), respectively, corresponding to rotate $|g|$ 
by $\pi/4$, $5\pi/4$, $7\pi/8$, and $13\pi/8$. It is noted that by comparing the biases derived from 
any two or three of the combinations to that from the four combinations, the results and conclusions 
are identical. Though with limited number of shear combinations, we conclude that it is sufficient to 
yield valid bias calibrations.

\begin{table}
\centering
\caption{The fractions of neighboring galaxies in the four CDFS tiles. 
Galaxies are defined as neighbors if their separation is less 
than $r$.}
\label{tab:blend}
\begin{tabular}{lrrr} 
\hline
Field & $r\leq1.0\arcsec$ & $r\leq2.0\arcsec$ & $r\leq3.0\arcsec$ \\
\hline
CDFS1 & 0.04\% &  4.83\%  & 16.34\% \\
CDFS2 & 0.06\% & 5.08\% & 16.73\% \\
CDFS3 & 0.03\% &   4.42\% & 15.99\% \\
CDFS4 & 0.05\% &   4.72\% &   16.52\% \\
\hline
\end{tabular}
\end{table}

\subsection{Simulated Images}\label{sec:simg}
Based on the mock catalog above, we generate as many single exposure images as the real observations. 
OmegaCam consists of $8\times4$ CCD chips, each one of $2047\times4000$ pixels with pixel scale of 0.214\arcsec.  
Our simulated single exposures have the same format.
To mimic the dither pattern we set the pointings of the simulated images to be exactly the same as in the observation. 
Because the imaging was conducted in many different nights, the background noise dispersion  
$\sigma_{\mathrm{bkg}}$ of the observed images after sky subtraction varies, typically ranging from 
10.0\,ADUs to 40.0\,ADUs with median of about 15.0\,ADUs. Such broad distribution contributes to different noise levels for a 
certain galaxy between different exposures, and hence can potentially bias the shape measurement. 
After applying masks in the observed sky-subtracted images, we find that the distributions of the pixel noise 
values in a single exposure can  be well described by a Gaussian function $N(0.0, \sigma^2_\mathrm{bkg})$. Therefore, the 
background noise of the simulated images is assumed to be Gaussian with $\sigma_{\mathrm{bkg}}$ 
fixed to that of the corresponding observation. To convert the apparent magnitudes to instrumental counts, 
the magnitude zeropoint is set to 24.58\,mag.

For each galaxy, \texttt{Galsim} can automatically assign a stamp size, and then 
project the surface brightness distribution onto the entire image stamp. The stamp size is 
typically larger than 30$\times$30 pixels, corresponding to several ten times of the scalelength, 
even for very faint galaxies. We point out that the fluxes in the parent sample are actually measured 
in a given aperture which is generally smaller than the total fluxes of 
galaxies \citep{1980ApJS...43..305K,1996A&AS..117..393B}. Similarly, if we again perform the 
same aperture photometry on the simulated images, the derived magnitudes from the stacked 
images will also be systematically fainter than the input, especially for those with large scalelengths, 
meaning that some faint galaxies in the parent sample cannot be 
detected after adding background noise. As a result, the magnitude distributions 
between the simulation and observation differ, especially at the faint end. To solve the issue, 
we shrink the stamp size of every galaxy based on its magnitude and half-light radius. Since 
\texttt{LensFit} truncates the model surface brightness distribution at a major-axis radius of 4.5 
exponential scalelengths for disc component or 4.5 half-light radii for bulge component, we fix the stamp 
sizes of galaxies fainter than 20.5\,mag to be 12.0 half-light radii, and 15.0 half-light radii for 
brighter ones, moderately larger than the model truncations in \texttt{LensFit}. With this adjustment, 
over 98\% of the input galaxies can be recovered in the final stacked image and the overall SNR distribution 
of them is consistent with observation as presented in Section \ref{subsec:vadSim}. 

We convolve the sheared galaxy profiles with the local PSFs, which are modeled by using the 
\texttt{PSFEx} package \citep{2011ASPC..442..435B} through observed single-exposure images. 
Observed stars with SNR  larger than 50 are selected for generating the PSF model used in the simulations here. 
A second order polynomial function is applied to model the variations over the entire CCD mosaic. 
Finally, the PSF at a given image position is calculated by a linear combination of six pixel basis vector images. 
The surface brightness profiles of the PSF-smeared galaxies and stars are then rendered onto the images.

For each shear combination, two sets of images are created where the galaxies in the second 
set are rotated by 90 degrees before applying shear and PSF convolution in order to reduce the 
shape noise \citep{2007MNRAS.376...13M}. The average of the intrinsic ellipticity is expected 
to be zero by this construction. In total, eight copies of each galaxy are simulated so that the total 
number of galaxies is about $4.8\times10^6$.

\subsection{Validation of the Simulation}\label{subsec:vadSim}

\begin{figure}
\centering
\includegraphics[width=0.23\textwidth]{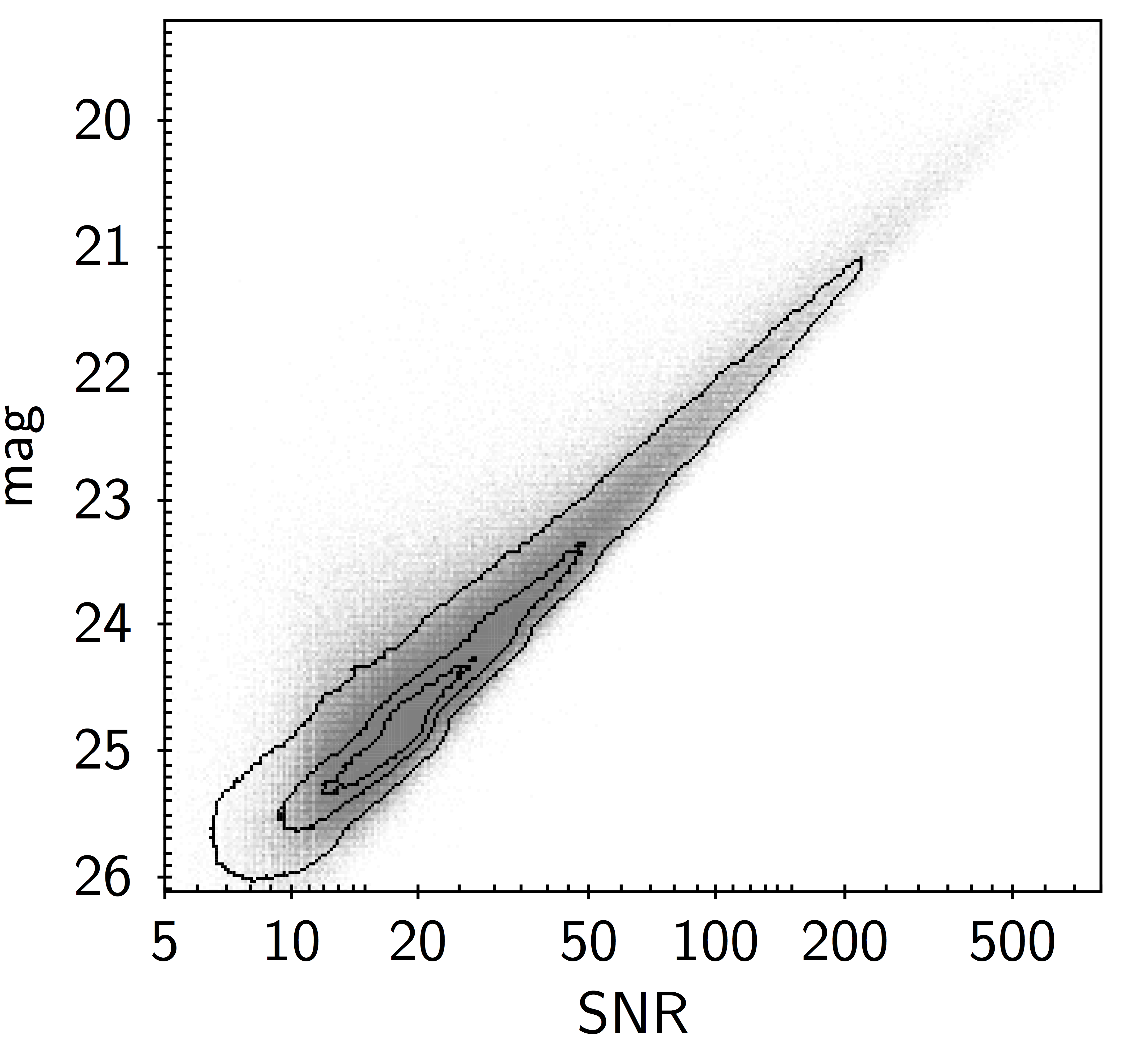}
\includegraphics[width=0.23\textwidth]{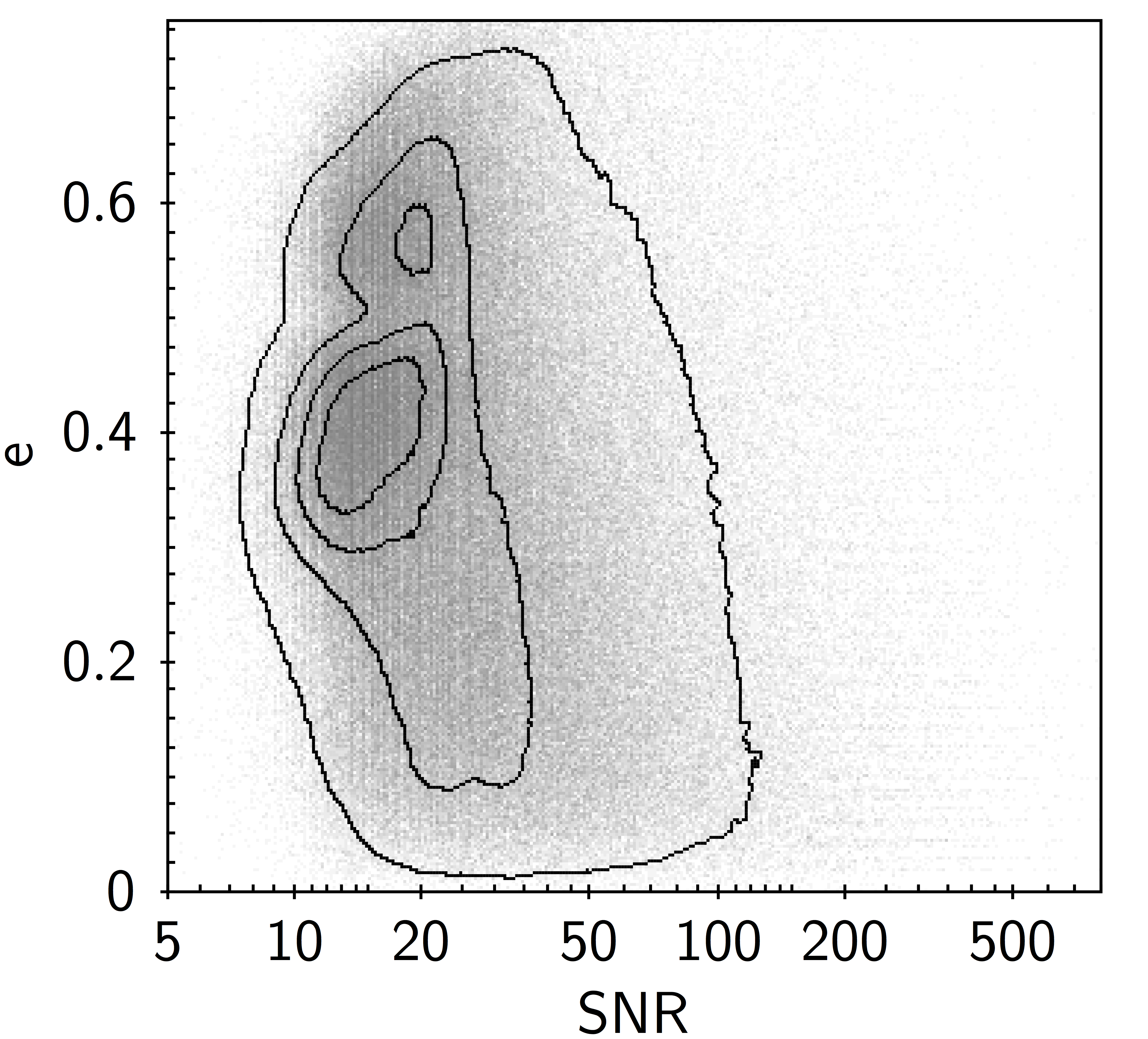}
\caption{The two-dimensional weighted  distributions of magnitudes and ellipticity versus SNR. 
The grayscale represents the data from VOICE observation, while the black contours are the 
density from simulation.}
\label{fig:bSNRcom}
\end{figure}

\begin{figure}
\centering
\includegraphics[width=0.45\textwidth]{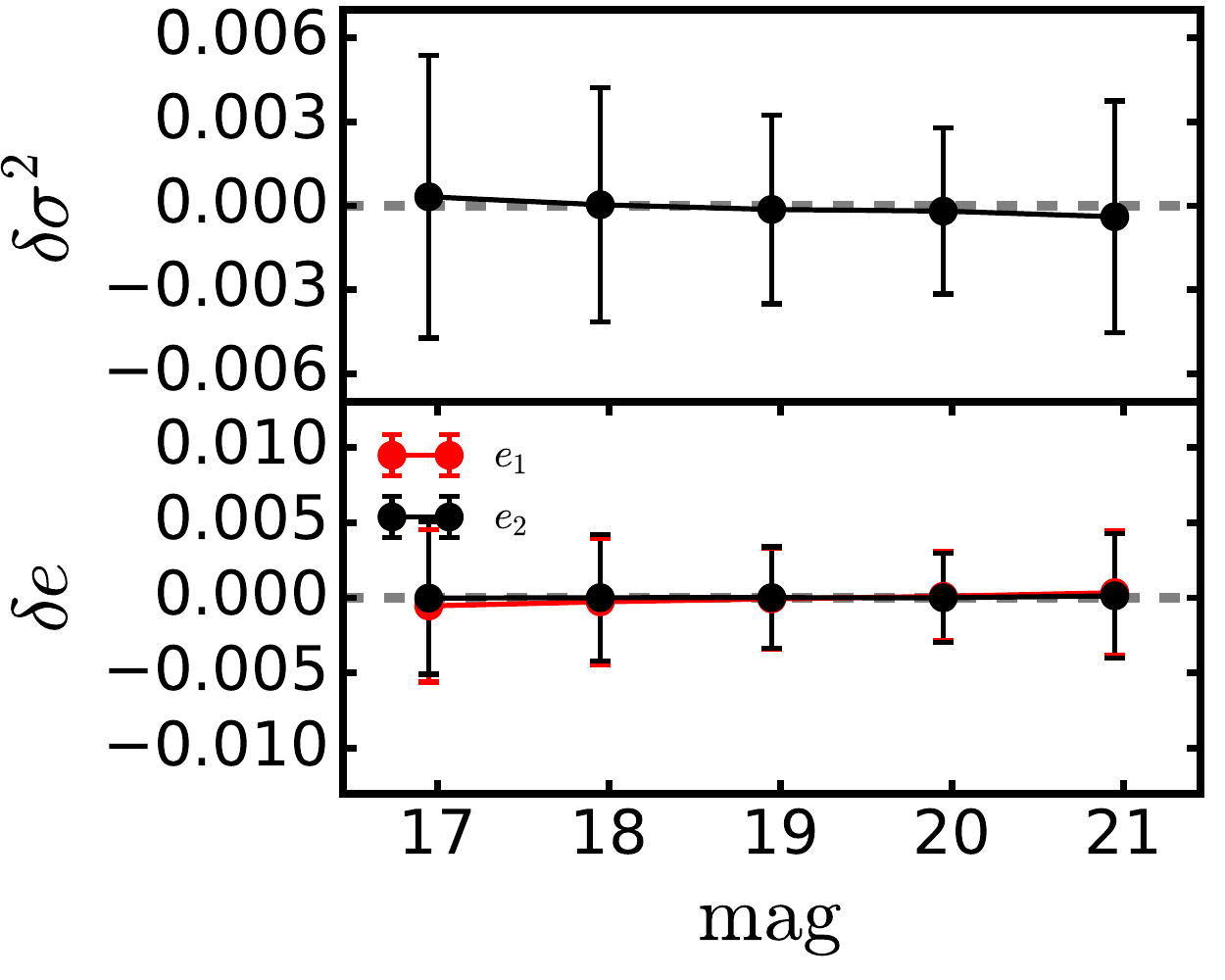}
\caption{Comparison between the size (top) and ellipticity (bottom) residuals of stars. 
These parameters are estimated directly from stars in the single exposure images and  PSF models  
constructed by \texttt{PSFEx}. The uncertainties are given by Poisson errors.}
\label{fig:PSFcom}
\end{figure}

\begin{figure}
\centering
\includegraphics[width=0.48\textwidth]{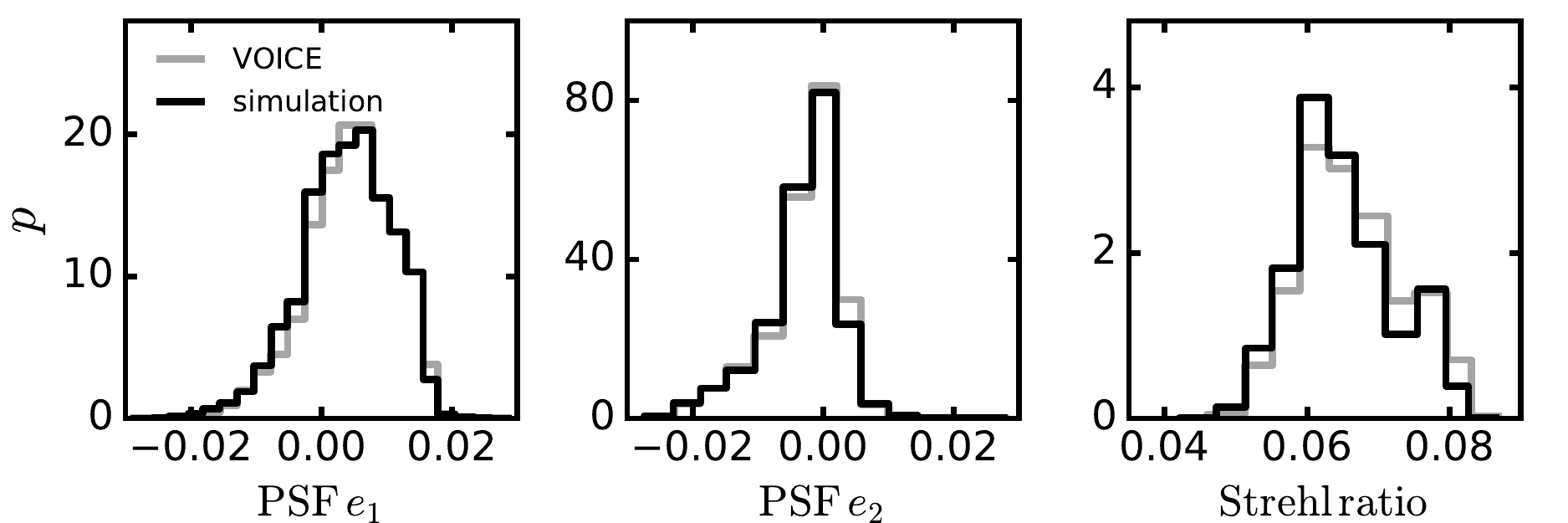}
\caption{Comparison of the weighted distributions of PSF parameters 
between the simulation (black lines) and VOICE observational data (grey lines).  
The distributions from left to right are the two ellipticity components ($e_{1}$ 
and $e_{2}$) and the Strehl ratio, respectively.}
\label{fig:comsopsf}
\end{figure}

\begin{figure}
\centering
\includegraphics[width=0.45\textwidth]{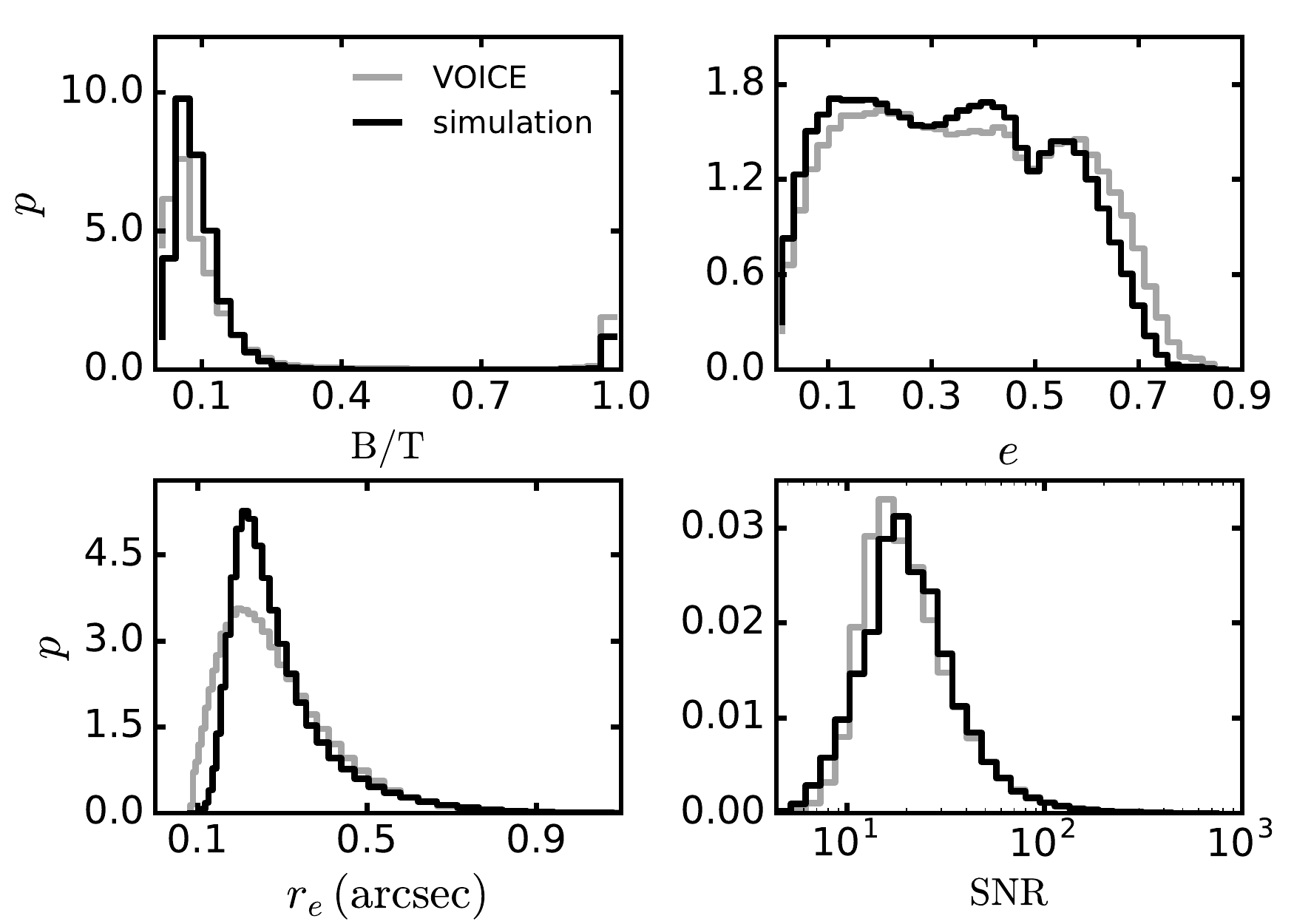}
\caption{Comparison of the weighted distributions of galaxy parameters 
from the simulation (black lines) and VOICE observational data (grey lines).  The distributions 
from left to right and top to bottom are the bulge fraction B/T, ellipticity, major-axis 
scalelength and SNR, respectively.}
\label{fig:comso}
\end{figure}

As described above, the simulated images correspond to the single exposure images 
of real observations after data reduction and astrometric and photometric calibrations. We note that we use the Gaia 
star catalog \citep{2016A&A...595A...2G} as reference to perform astrometric calibration on the real 
images, and the positional dispersion is 0.056\arcsec (see F18). With such a sub-pixel accuracy, the astrometric 
residuals do not contribute significant systematics to the measured shear signals. We thus do not include the 
uncertainties in the astrometric calibration in our simulations.
The dither pattern and gaps between CCD chips 
are set to be the same as in the real observations.  In this section, we validate quantitatively our simulations 
by comparing to the real observations.

We note that although the global background noise levels of the simulated images are identical to those of observations, 
the local variations are not necessarily the same. As a result, the positions and magnitudes of the objects 
derived from the simulations do not exactly match the input values. This can affect the source detection. 
For the simulations to be self-consistent, we therefore firstly stack the simulated 
images and re-perform source detection and photometry using \texttt{SExtractor} with the same parameters 
as in the production of the photometric catalog from the data. These new source catalogs are used as input for 
shape measurements on the simulated images. 
Then we follow the same procedures as in real observational analyses described in 
Section~\ref{sec:ShearMeasure} to measure the shapes of galaxies and to obtain the 
shear catalogs for the eight sets of simulated images.

Several cuts are then applied to the shear catalogs for bias analysis. First of all, only galaxies with 
non-zero weight are selected. Since the properties of galaxies are analyzed as weighted average 
in the following sections, this cut does not affect our bias calibrations. We further reject the potentially 
problematic galaxies which are flagged as non-zero by \texttt{LensFit}. After these 
constraints, we match these catalogs with the mock sample described in Section~\ref{sec:mcat} to 
obtain the true shear values for the galaxies. This matching is done using a $k$-d tree nearest neighbor search 
algorithm which is fast and efficient for large dataset. An appropriate aperture selection for the matching is 
essential given the existence of neighboring galaxies. A larger aperture can increase the probability of 
spurious matches, while too small aperture makes many faint galaxies miss out due to the noise-induced mismatch 
of the coordinates.  It is found that an aperture radius of 0.6 arcsec, corresponding 
to three pixels, can efficiently remove spurious detections and reduce the probability of mismatch for neighboring galaxies.
Finally, about 2.3 million galaxies are obtained in the sample for bias analysis. 
Figure \ref{fig:bSNRcom} shows the two-dimensional weighted distributions of magnitudes and ellipticity 
versus SNR for the observation and simulation data. The double-peak distribution of ellipticity is attributed to  different 
ellipticity priors between the disc-dominated and bulge-dominated galaxies \citep{2013MNRAS.429.2858M}.  

Since the dominant contribution to the shear biases results from the imperfect PSF modeling, 
an appropriate simulation should be capable to capture the main features of the observed PSF,
especially the spatial and temporal variations. 
To validate the PSF model used in the simulations, we follow a similar 
methodology of \citet{2017arXiv170801533Z}. We calculate the size 
and shape residuals between the stars in the observed single exposure images and the corresponding PSF 
modeled by \texttt{PSFEx}. The stars are identified by finding the stellar locus in the size-magnitude 
diagram with magnitude $16.0<r_{\mathrm{mag}}<22.0$ and SNR higher than 20. Compared to the stars 
used for PSF modeling as described in Section~\ref{sec:simg}, we identified more fainter and lower SNR stars for testing. 
The size and the shape are estimated adaptively by calculating the moments of the light 
profile \citep{2003MNRAS.343..459H}, encoded 
in the \texttt{Galsim} toolkit. This method can estimate the best-fit elliptical Gaussian to the star and calculate the $\sigma$ value 
(in unit of arcsec; defined as $|\det(M)|^{1/4}$ where $M$ is the metrix of the moments) as a representation of the size. 
The ellipticity is defined as $\bf{e} = (a-b)/(a+b)\exp(2\mathrm{i}\theta)$ where $a$, $b$ and $\theta$ are 
the major axis, the minor axis, and the orientation of the best-fit ellipse, respectively. Figure~\ref{fig:PSFcom} compares the 
size and ellipticity residuals of the observed stars and the modeled PSF interpolated to the same image positions. The dots 
indicate the median residuals in each magnitude bin, while the uncertainties are given by Poisson errors. 
Within the errorbars, the  size and ellipticity residuals are consistent with zero within the magnitude range, showing a 
good agreement between the modeled PSF used in our simulations and that of real observations. 

Further comparisons of the PSF parameters estimated by \texttt{LensFit} between the simulation and 
observation are shown in Figure~\ref{fig:comsopsf}. It can be seen that the weighted distributions of the two ellipticity 
components and the Strehl ratio\footnote{The Strehl ratio is generally defined as the ratio of the peak 
aberrated intensity relative to the maximum attainable intensity from a point source in diffraction-limited 
optical system. In \texttt{LensFit}, it is defined as the fraction of PSF light contained in the central pixel.}  
parameter measured from simulated images are in good agreement with the observed data. The significant 
difference in distributions between the two 
ellipticity components implies the complicated PSF variations in the observations. One possible 
reason is that the long time span in observations for every tile makes the PSF pattern varied 
remarkably. The small survey area of VOICE may also be a reason because certain differences of 
$e_1$ and $e_2$ can persist. This is in contrast to surveys with a large sky coverage, for which the 
statistical distribution of PSF over all the fields is approximately isotropic. 
Our PSF models can properly reproduce the PSF features existed in VOICE observations.

In addition, Figure~\ref{fig:comso} further compares the weighted distributions of some 
galaxy parameters measured by \texttt{LensFit} from the simulation and observed data. The bulge 
fraction derived from simulation is well-matched with the VOICE data. The small differences of the 
scalelength and SNR indicate that small and faint objects are still absent in our simulation although 
we have reduced the stamp size in simulation according to the galaxies' scalelength and magnitude 
to suppress this effect. The discrepancy presumably results partly from the different intrinsic size 
distributions between the simulation and real observation, and the fixing of background noise dispersions 
in the simulation. We can also see small differences in the ellipticity distributions.   Such mismatch may 
indicate that the intrinsic ellipticity distribution used in the simulation is not exactly the same as that in the 
real observation.  However, as demonstrated in the simulation of KiDS survey \citep{2017MNRAS.467.1627F} 
where similar discrepancies in the distributions of size, SNR and ellipticities  
are presented,  the resulting biases for the shear calibration are negligible.  Through changing the ellipticity distributions, 
similar conclusion was also drawn even in the \textit{Euclid}-like simulation \citep{2017MNRAS.468.3295H}.
We will discuss these more in Section~\ref{sec:dis}.

\section{Bias Calibration}\label{sec:BiasCal}
Following \citet{2006MNRAS.368.1323H}, the accuracy of the reduced shear 
$g^\mathrm{obs}$ can be modeled in terms of the multiplicative bias $m$ and 
additive bias $c$ relating to the true shear $g^\mathrm{true}$ as

\begin{eqnarray}
g^\mathrm{obs}_{i} = (1+m_{i}) \times g^\mathrm{true}_{i} + c_{i}, \nonumber
\end{eqnarray}
where $g^\mathrm{obs}$ denotes the weighted average of the ellipticity measured by 
\texttt{LensFit} and the subscript $i$ refers to the two shear components.

The multiplicative bias and additive bias generally depend on the observed galaxy 
properties, such as the SNR and galaxy size. The additive bias primarily stems from the 
residuals in modeling the PSF anisotropy. It can be empirically corrected using the 
observed data. The multiplicative bias, a change of the amplitude of the 
shear, is mainly attributed to the background noise and pixelation, and most likely 
affects the shape estimate of faint galaxies. It is generally calibrated through image 
simulations. 
 
In this section, we perform detailed bias calibrations for the measured galaxy shape. 
We note that the binning strategy for each observable adopted in this work is by 
equalizing the total number of galaxies in each bin. The weighted average is assigned 
as the center of each bin for the corresponding observable. We use the bootstrap 
method with 100 realizations to derive the uncertainties of the estimated shear in 
each bin. The $\chi^2$ minimization is then applied to yield the multiplicative and 
additive biases as well as the associated uncertainties. 

\subsection{Selection Bias}
Besides the bias resulting from noise and model fitting, the source detection and shape 
measurement procedures can also introduce bias. This kind of bias is usually referred 
as selection bias and it was extensively discussed in many studies through image 
simulations \citep{2000ApJ...537..555K,2002AJ....123..583B,2003MNRAS.343..459H,
2006MNRAS.368.1323H}. Due to the difficulties in accurately measuring the shape of 
faint and small galaxies for many shape measurement methods,  these galaxies suffer 
from more severe selection bias than the bright ones. Therefore, the bias is expected to 
be a function of the magnitude (or equivalently the SNR) and galaxy size, and can arise 
in both the observation and simulation. In the KiDS simulation, \citet{2017MNRAS.467.1627F} 
reported a significant multiplicative selection bias which was as large as 4.4\% after 
averaging the true sheared ellipticity of galaxies with non-zero weights. It showed an obvious 
dependence on the magnitude and major-axis scalelength $r_{e}$. However, the dependency  
is reduced when we consider the geometric average of the major- and minor-axis scalelengths, 
denoted as $r_{ab}$, because it is less correlated with the measured ellipticity.

\begin{figure}
\centering
\includegraphics[width=0.45\textwidth]{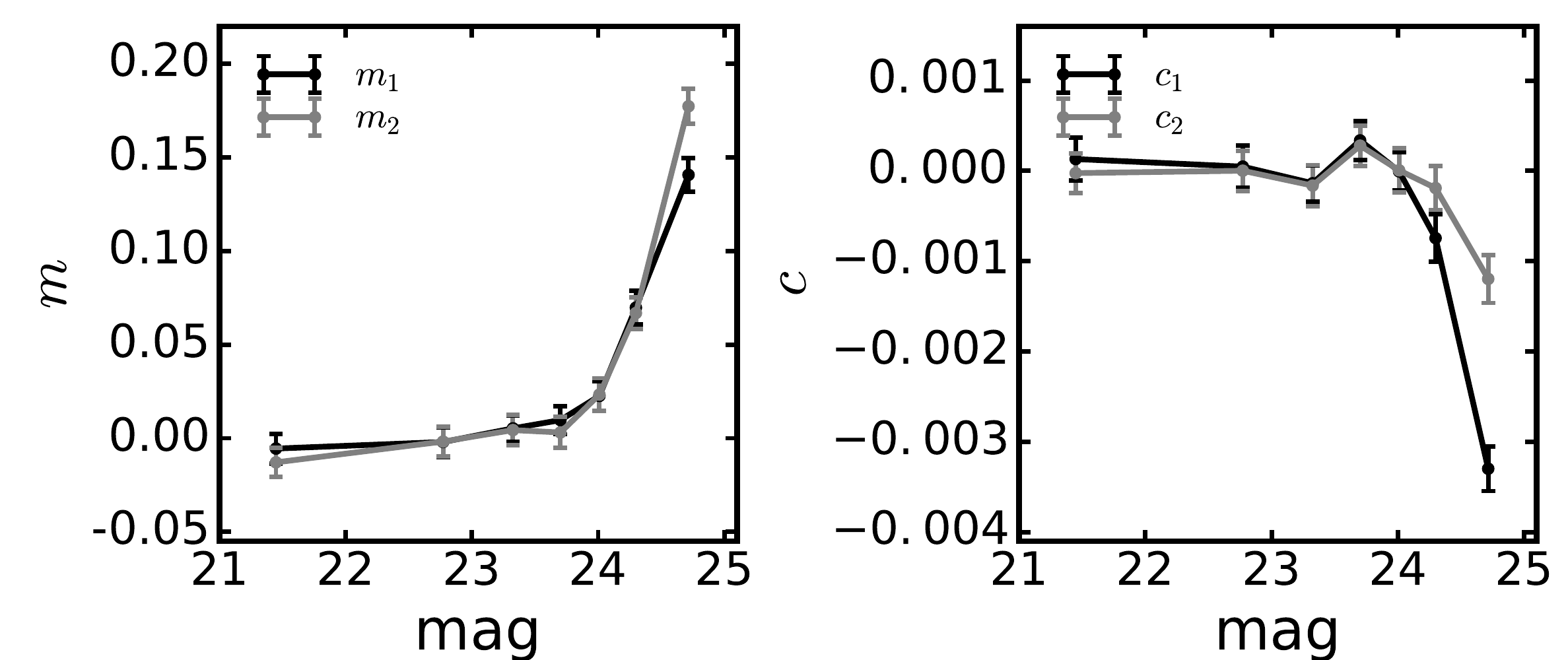}
\includegraphics[width=0.45\textwidth]{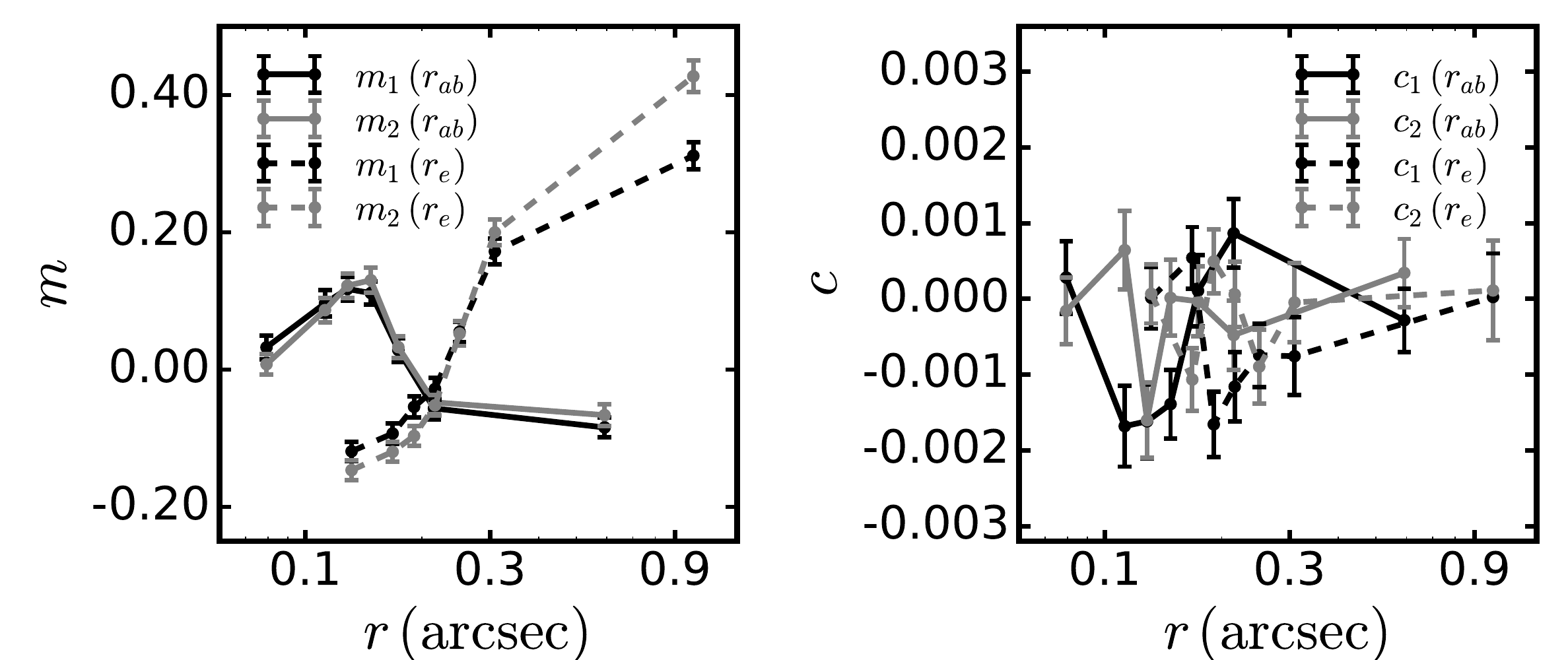}
\caption{\textit{Top panel}: the multiplicative and additive selection biases as a function of magnitude 
for galaxies with non-zero weight. \textit{Bottom panel}: the multiplicative and additive selection biases 
as a function of galaxy size. The solid lines represent the size $r_{ab}$ defined as the geometric 
average of the major- and minor-axis scalelengths, while the dashed lines indicate the 
scalelength $r_{e}$ along major axis. Note that both of them are calculated through \texttt{LensFit} 
output.}
\label{fig:bselect}
\end{figure}

Following a similar scheme, in our simulation,  we quantify the selection bias by analyzing the input sheared ellipticity 
of galaxies in terms of the observables. 
As described in Section~\ref{sec:ImgSim}, only galaxies 
detected in the observations are used to generate the simulated images, and we find similar 
number of galaxies with shape measurements compared to that of observation. For those galaxies 
detected in the simulated mosaic images using \texttt{SExtractor}, the selection bias is derived by comparing the  
average of the input true sheared ellipticity with true shear. In this case, the effect of noise bias vanishes, and 
biases stemming from the detection procedures, including the potential cancellation of zero shape noise 
implementation due to undetected galaxies, are dominant.
It turns out that the selection bias is almost negligible at this stage.  
On the other hand, if considering only the galaxies with non-zero weight after running \texttt{LensFit}, 
the multiplicative selection bias becomes apparent for faint galaxies, as shown in Figure~\ref{fig:bselect}.  The top panel of Figure~\ref{fig:bselect} 
displays the dependence of the selection bias on magnitude. It can be seen that the multiplicative selection 
bias is nearly zero at magnitude brighter than 24.0\,mag. It increases dramatically at fainter magnitudes because of the 
noise effect that results in a considerable  fraction of shape measurement failure.  Similar trend can also be seen for the additive bias. 
The multiplicative bias also exhibits a strong dependence on the major-axis scalelength $r_{e}$, 
as shown in the bottom panel of Figure~\ref{fig:bselect}. However, an apparent lower correlation 
is seen by adopting $r_{ab}$. The additive bias does not present significant correlations 
with either definition of the galaxy size. In our following analyses, we use $r_{ab}$ as the proxy 
of galaxy size to perform bias calibration.

As discussed above, the shear signals measured from both simulation and 
observation are supposed to be subjected to the selection bias.  Therefore, to calibrate the 
shear in the VOICE survey, it is essential to take into account all the sources of bias, including the 
selection bias, the noise and model biases, through our simulation. In the following sections, 
we systematically investigate the bias calibration based on the observables SNR and $r_{ab}$ 
since they are the two predominant quantities that the bias depends on.

\subsection{Empirical Calibration}

\begin{figure}
\centering
\includegraphics[width=0.45\textwidth]{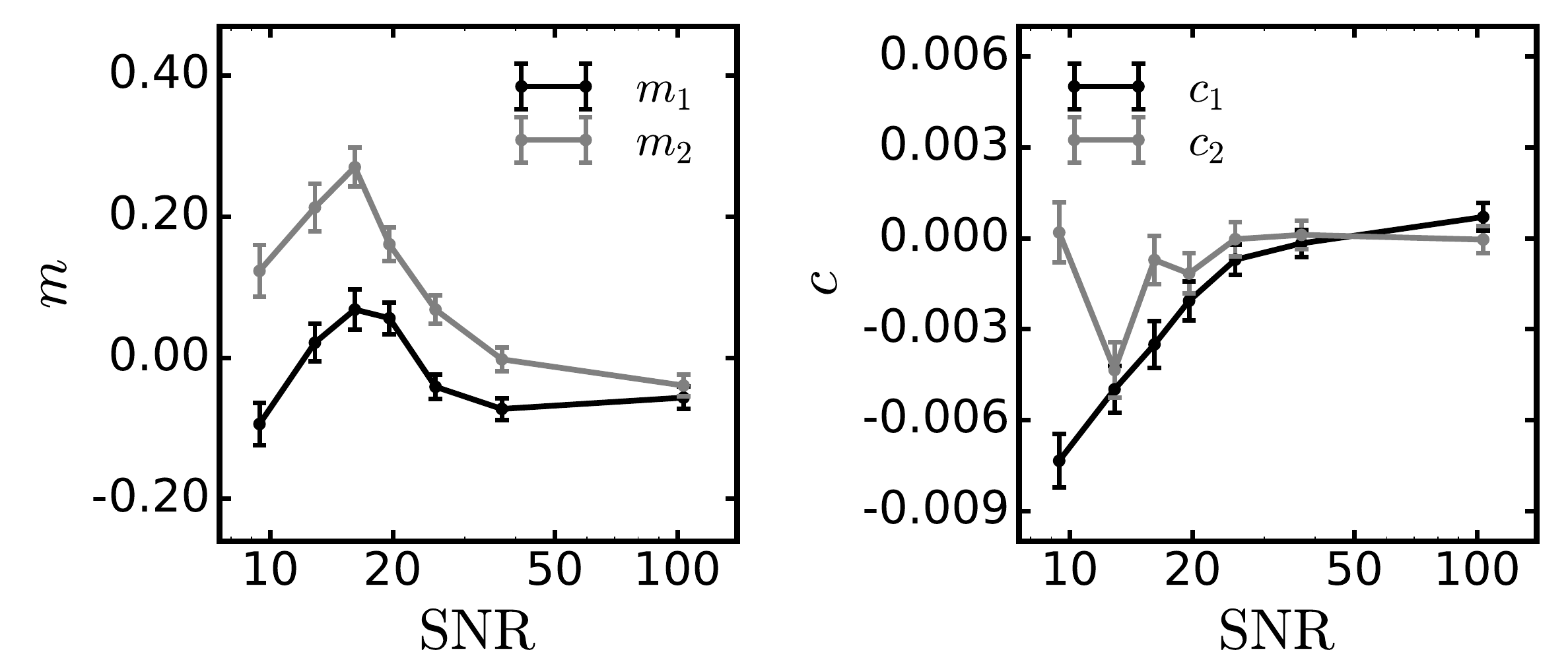}
\includegraphics[width=0.45\textwidth]{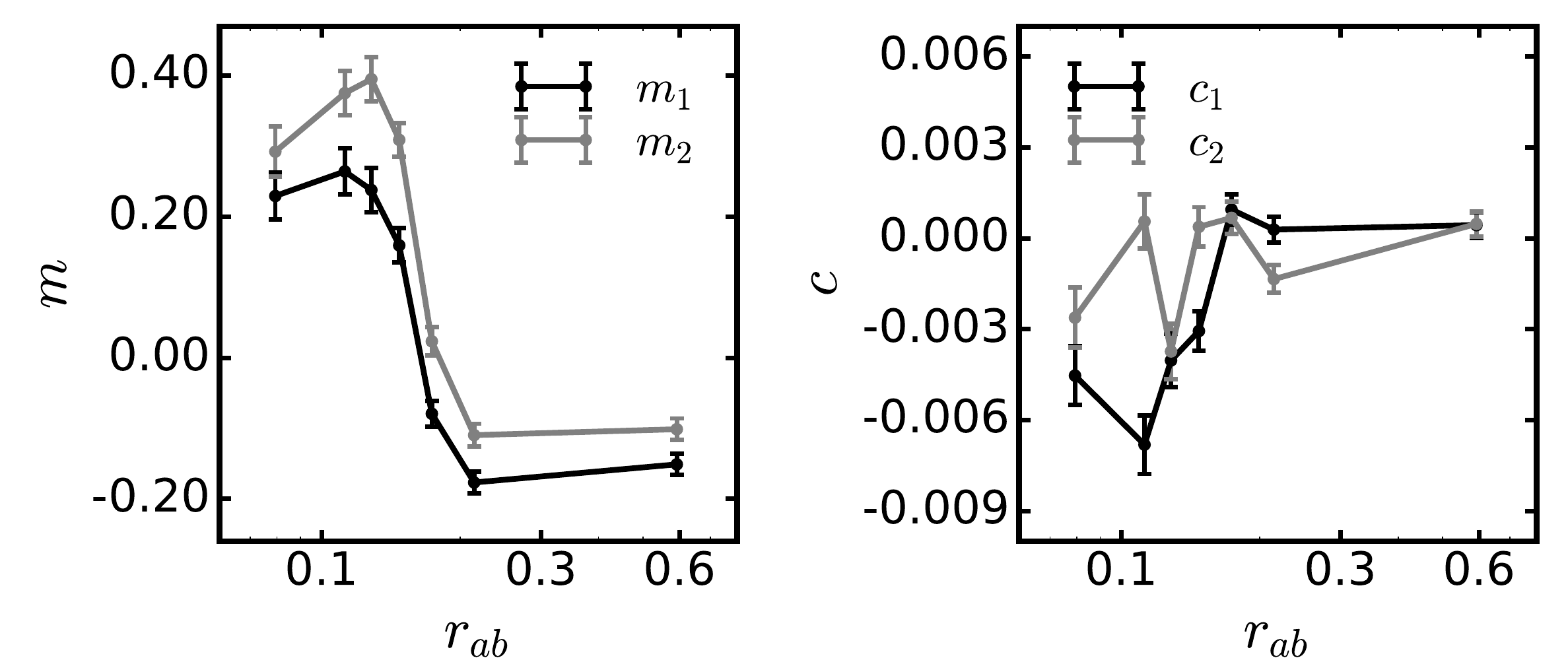}
\caption{The multiplicative and additive biases of the measured shapes as a function of 
SNR (\textit{top panel}) and size (\textit{bottom panel}).}
\label{fig:biasAna}
\end{figure}

\begin{figure}
\centering
\includegraphics[width=0.23\textwidth]{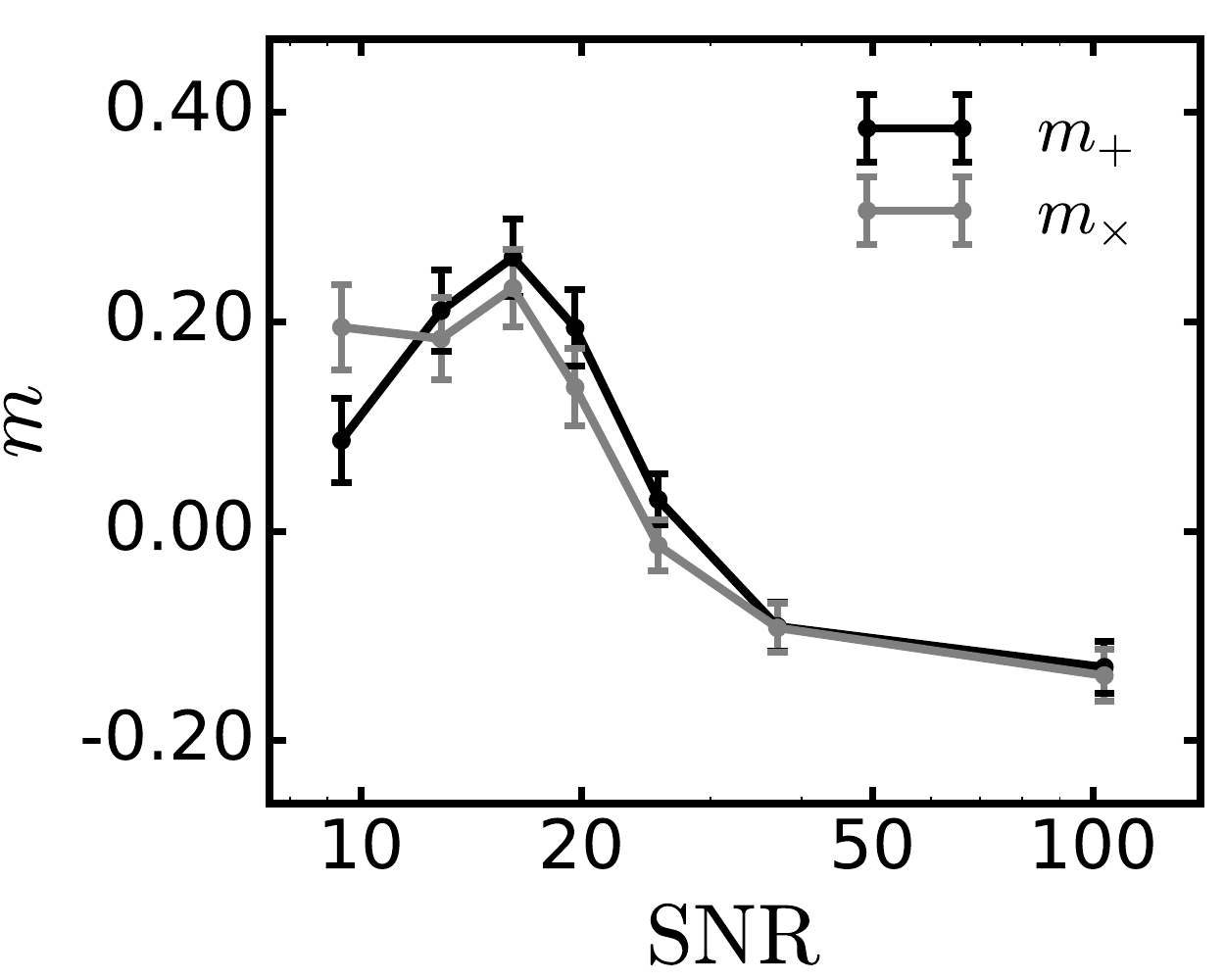}
\includegraphics[width=0.23\textwidth]{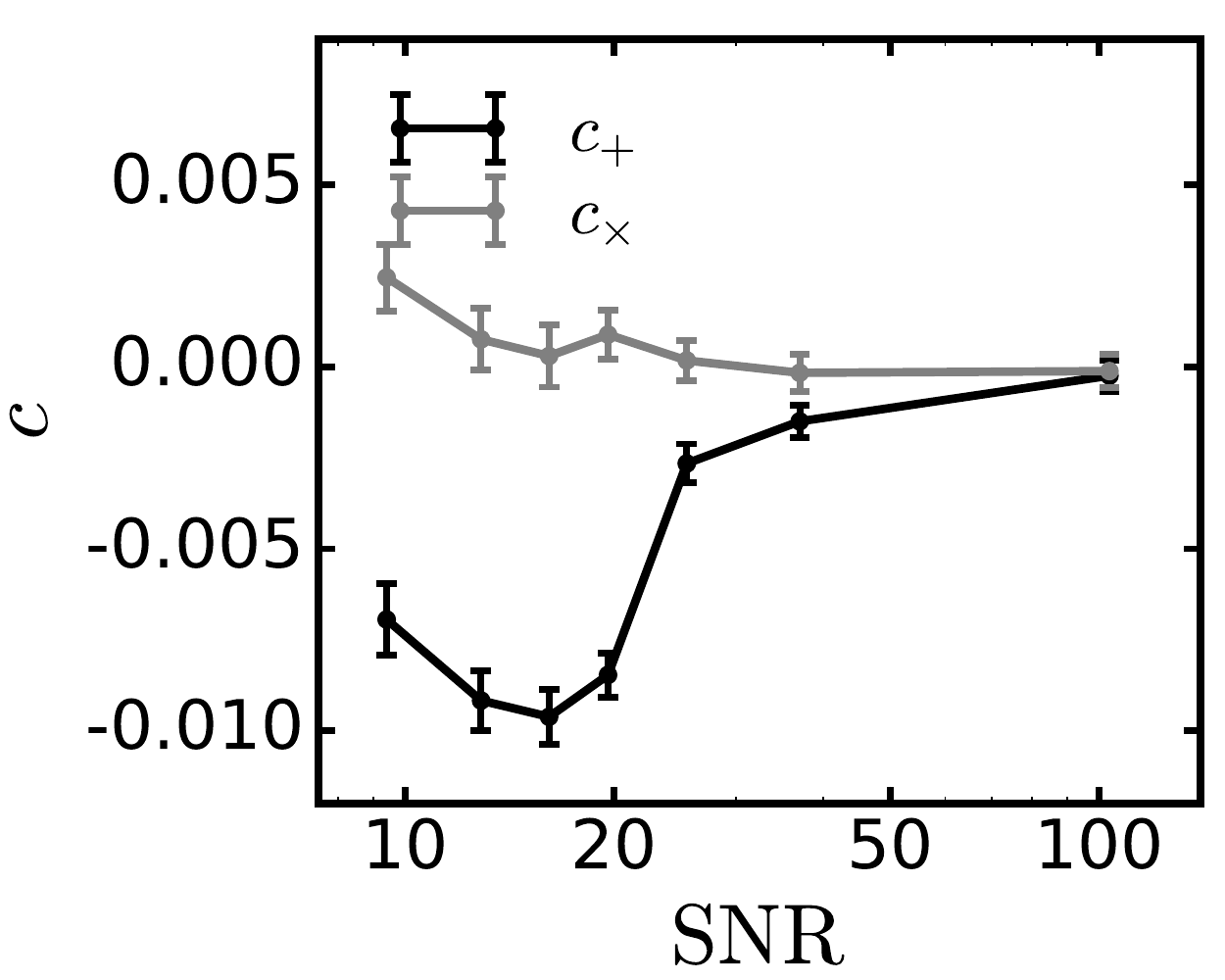}
\caption{The multiplicative ($m_{+}$, $m_{\times}$) and additive 
($c_{+}$, $c_{\times}$) biases of the measured shapes as a function of 
SNR. The bias components are derived by aligning the galaxy's ellipticity and shear to 
the corresponding PSF ellipticity.}
\label{fig:biasPSFAlign}
\end{figure}

\begin{figure}
\centering
\includegraphics[width=0.48\textwidth]{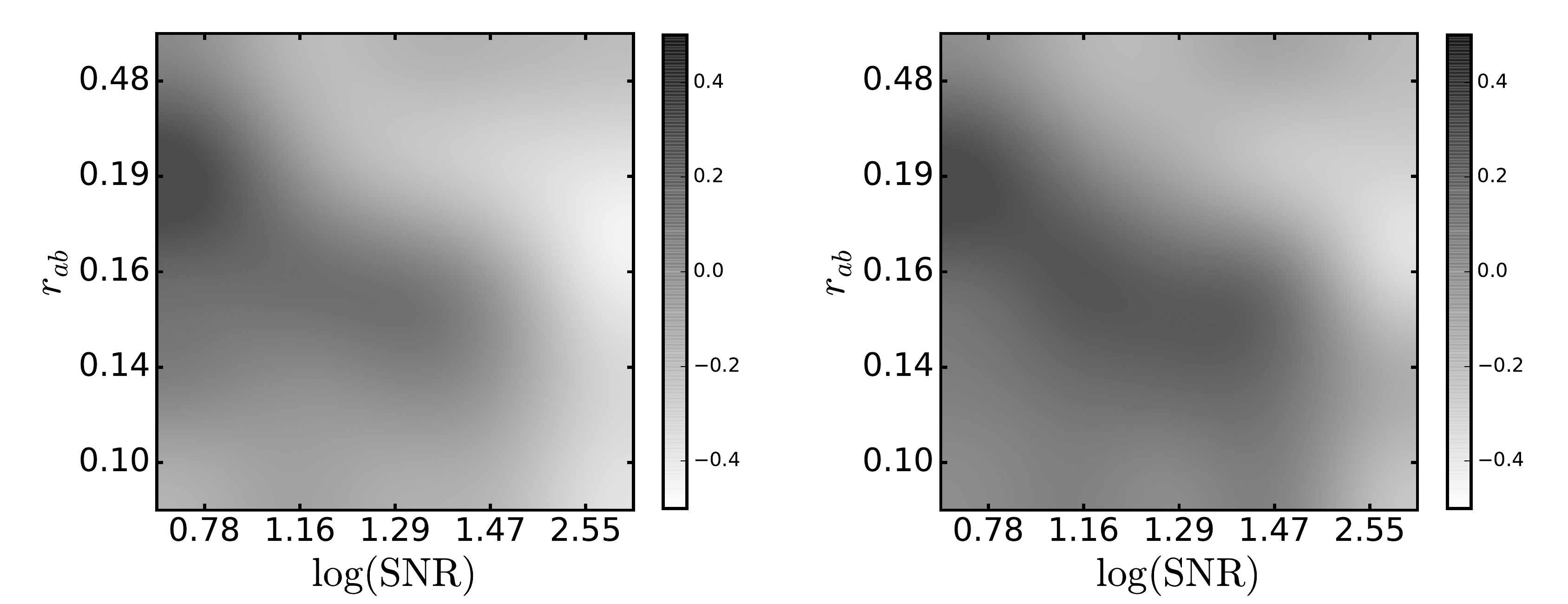}
\includegraphics[width=0.48\textwidth]{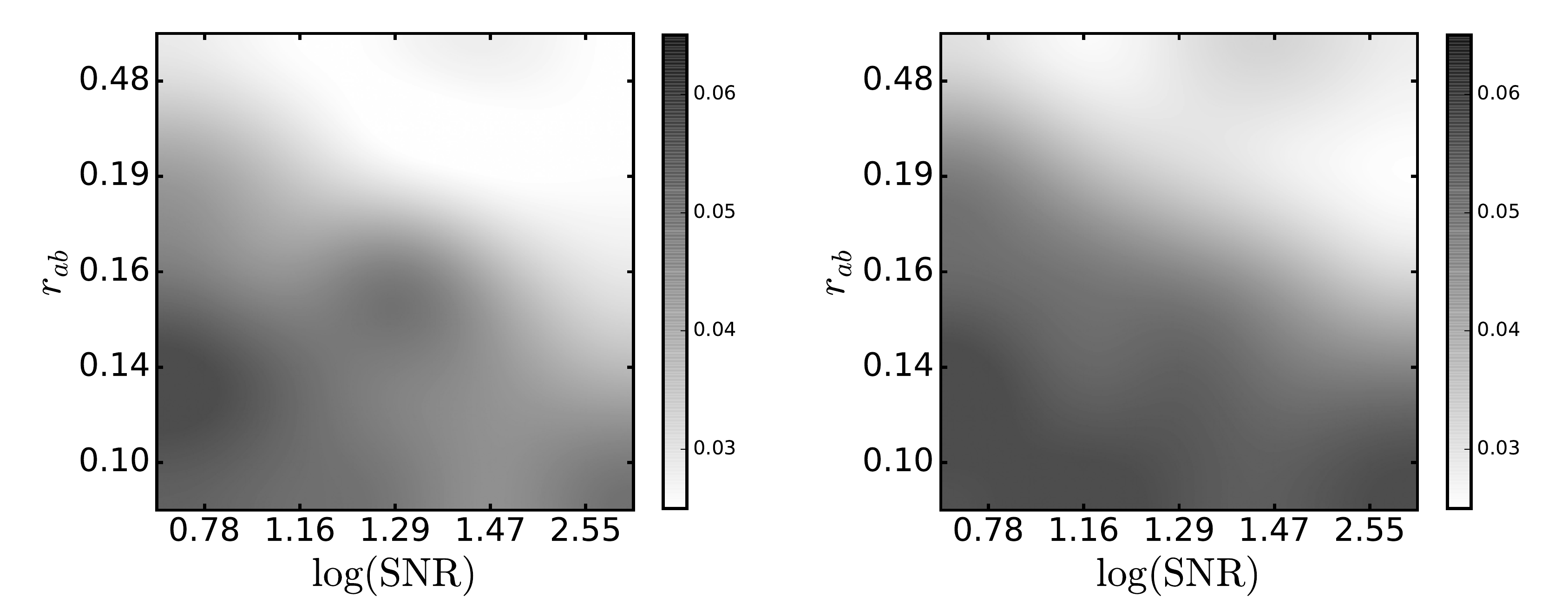}
\caption{\textit{Top panel}: Multiplicative bias distributions ($m_1$ on the \textit{left} and $m_2$ on the \textit{right}) 
in the SNR\,--\,$r_{ab}$ two-dimensional plane. For clarity, a ``lanczos" kernel is applied to smooth the discrete values.
\textit{Bottom panel}: Estimated error distributions of the multiplicative biases.}
\label{fig:biasSurf}
\end{figure}

The biases of the shear measured from the simulation as a function of galaxy 
SNR and size are shown in Figure~\ref{fig:biasAna}. As expected, both the 
multiplicative bias and additive bias get larger for galaxies with low SNR and 
small size. The maximum of the absolute values reaches 0.4 for multiplicative 
bias and 0.008 for additive bias. One feature shown in Figure~\ref{fig:biasAna} is that the 
two components for both multiplicative and additive biases present somewhat different amplitudes.
As discussed in \citet{2007MNRAS.376...13M} and \citet{2015MNRAS.450.2963M}, 
the additive bias components in the pixel coordinate frame probably result from the selection 
bias and potential numerical artifacts.  The difference between the two multiplicative components $m_1$ and $m_2$ 
may be due to the effect of pixelization of the galaxy images and PSF profiles.
In order to check the hypothesis, we compute the tangential and cross components of the shape and 
shear for each galaxy in a reference system aligned with the PSF ellipticity axes. The derived bias components are 
defined as ($m_{+}$, $c_{+}$) and ($m_{\times}$, $c_{\times}$), respectively. 
Since the PSF ellipticity is approximately randomly orientated with respect to the pixel axes, the difference 
due to pixelization is expected to be cancelled out. As illustrated in 
Figure~\ref{fig:biasPSFAlign}, the $m_{+}$ and $m_{\times}$ have much more similar amplitudes than 
that of $m_1$ and $m_2$. The small residual difference might be related to the somewhat 
different distributions of the two PSF components shown in Figure~\ref{fig:comsopsf}. The PSF anisotropy explains the difference 
of $c_{+}$ and $c_{\times}$. To simplify the shear analysis, in the following we will focus on the shear calibration in the original pixel frame.  

Because of the amplitude differences for the two components, we cannot adopt a uniform analytical 
expression, such as the function used in \citet{2013MNRAS.429.2858M}, to describe the calibration parameters. 
We therefore take the similar approach applied in KiDS simulation \citep{2017MNRAS.467.1627F} to 
use the bin-matching method on the SNR\,--\,$r_{ab}$ surface to calibrate the bias. Specifically, 
we firstly bin the galaxies by SNR and size $r_{ab}$ in the two-dimensional plane, and then 
derive a constant bias in each bin. If one observed galaxy falls into a certain bin, its ellipticity 
will be calibrated by applying the corresponding biases. 

\begin{figure}
\centering
\includegraphics[width=0.45\textwidth]{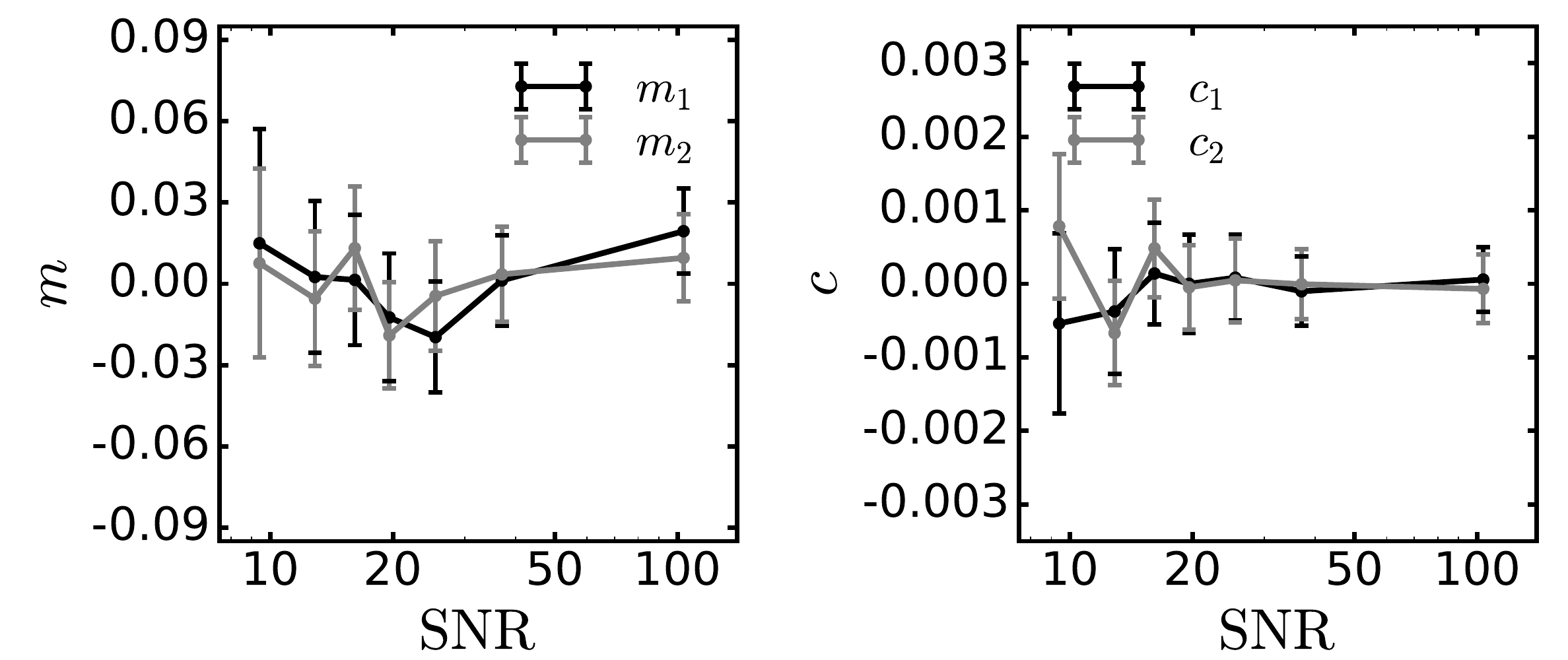}
\includegraphics[width=0.45\textwidth]{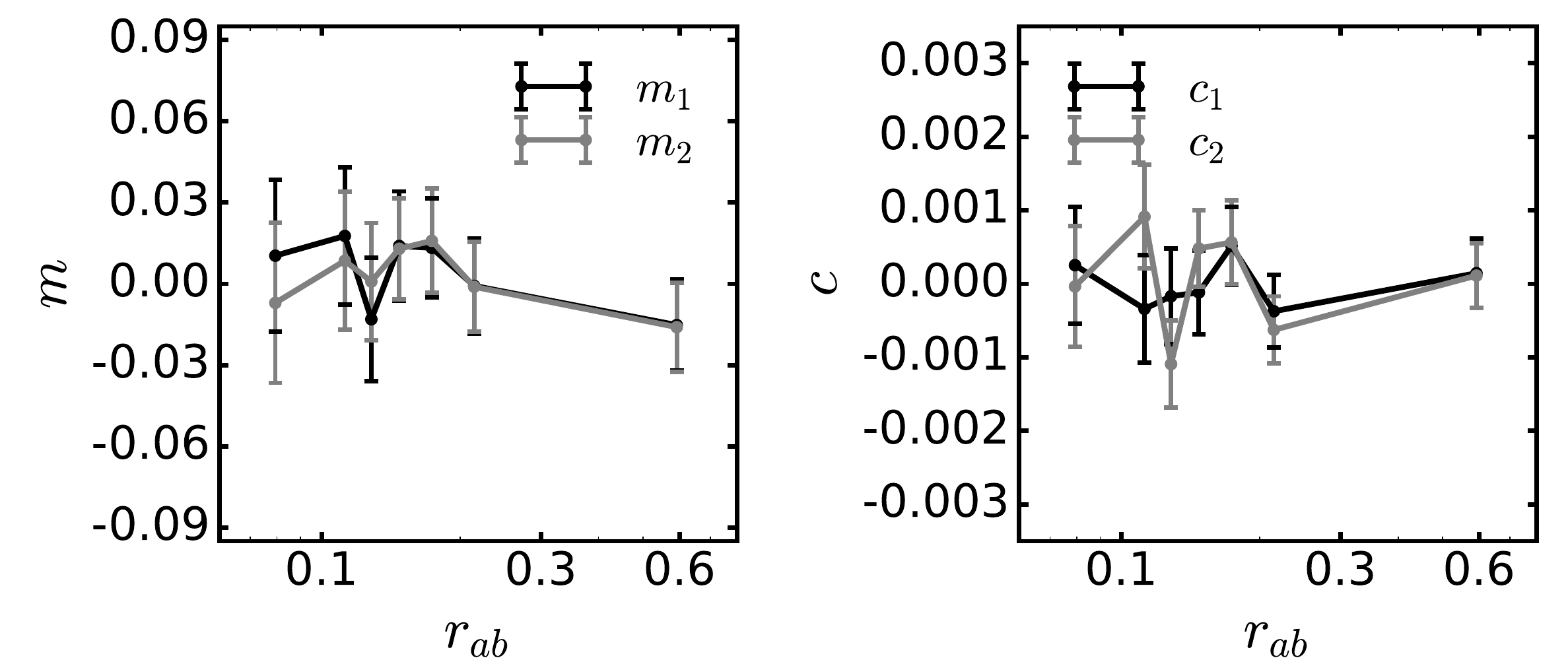}
\caption{The residual multiplicative and additive biases after calibration using the bin-matching 
method as a function of measured SNR (\textit{top panel}) and size (\textit{bottom panel}).}
\label{fig:biasCal}
\end{figure}

Due to our limited sky coverage and relatively small amount of galaxies, an appropriate 
binning scheme is crucial to derive valid bias calibration results. If the 
number of bins is too small, we may miss out on some real features in the bias 
surface. However, the statistical uncertainty arises if there are too many bins, and 
that can result in extra artificial bias. For our simulation, we find that a five-bin scheme 
along both SNR and $r_{ab}$ axis can yield robust calibration. In this case, the average 
error of the multiplicative bias in each bin is 0.04, while the SNR$_{m}$ (defined as $m/m_{err}$, 
where $m_{err}$ is the estimated error of $m$) is close to 4.8. The top panel of Figure \ref{fig:biasSurf}
illustrates the two-dimensional distributions of the two multiplicative bias components in the SNR\,--\,$r_{ab}$ 
plane. It is seen that while they present similar dependence on the two observables, the amplitudes of $m_2$ are 
systematically larger. The distributions of corresponding error $m_{err}$ are shown in the bottom panel. As expected, 
galaxies with smaller size and lower SNR suffer more significant calibration uncertainties. 
Figure~\ref{fig:biasCal} shows the final residuals after bias calibration. It can be seen that the multiplicative bias is 
well within 0.03 over the entire SNR and size ranges, and the additive bias almost vanishes. The residuals 
do not present strong dependence on the SNR and galaxy size. Overall,  both the residual multiplicative bias and 
additive bias are consistent with zero, indicating that the calibration is unbiased.

\section{Calibration Sensitivity}\label{sec:dis}
As we have discussed in Section~\ref{sec:ImgSim}, the sample extracted from simulation misses some  
faint and small size galaxies. The distributions of the ellipticity between the simulation and 
observation are also slightly different. These differences may result in extra residual bias when 
applying the calibration results to observation. \citet{2017MNRAS.468.3295H} studied the 
sensitivities  to these effects based on \textit{Euclid}-like image simulation, and concluded 
that the multiplicative bias is indeed affected by these factors. However, they demonstrated that 
the amplitude change of the multiplicative bias is always less than 0.005 by 
varying the corresponding distributions. Similar conclusions were also drawn in the KiDS 
simulation \citep{2017MNRAS.467.1627F}, which stated that the sensitivities of the multiplicative bias to the different distributions 
can be safely neglected for the present accuracy requirement ($m\sim0.01$) in weak lensing 
surveys. VOICE and KiDS surveys share the same instrument and observational 
configuration. The VOICE survey is deeper, but the area is much smaller than that of KiDS. 
Thus the number of galaxies with successful shear measurements is smaller, resulting in larger 
statistical uncertainties in cosmological analyses. We therefore expect that the effect of lacking 
of small and faint galaxies in our simulation on the shear bias calibration is even less significant 
than that of KiDS. However, for future deep and wide surveys, this can be an issue \citep{2017MNRAS.468.3295H}.

On the other hand, since the galaxies in our simulation are only from observation without 
including those below detection limit, the undetected galaxies may introduce potential 
bias. In addition, since the positions of galaxies in the simulated images exactly match those of real galaxies, it is possible 
to study the impact of the galaxy blending effect on the multiplicative bias. We focus on 
the sensitivity analyses of these two factors in this section.

\subsection{Impact of Galaxies Below Detection Limit}

\begin{figure}
\centering
\includegraphics[width=0.45\textwidth]{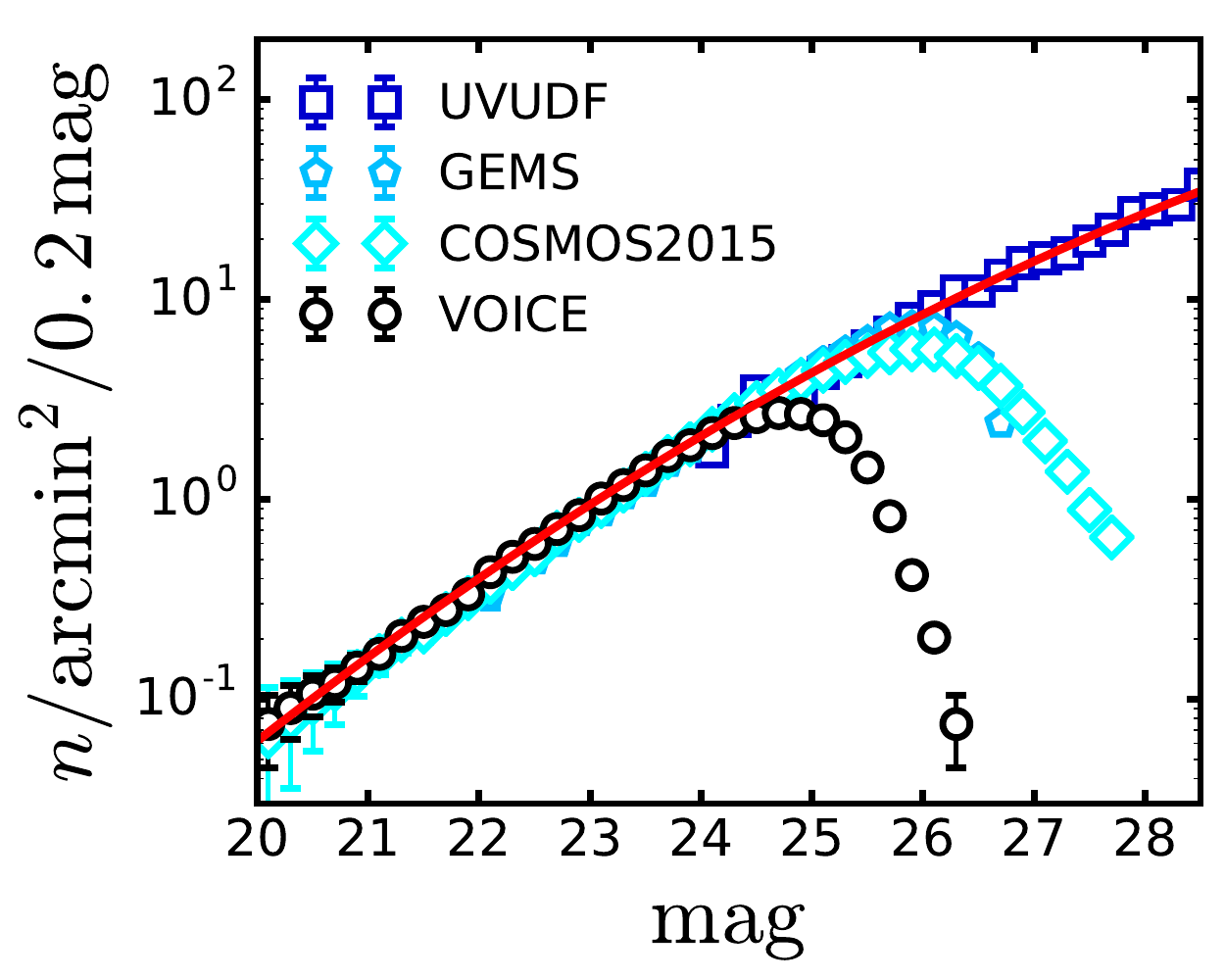}
\caption{Magnitude distributions of different $r$-band photometric catalogs with errorbar estimated 
by the Poisson statistics. The magnitudes are from UVUDF F606W band (blue), 
GEMS F606W band (light blue), COSMOS2015 $r$ 
band (cyan) and VOICE $r$ band (black), respectively. The red solid line 
represents the best fit for VOICE $20.0<m_r<24.0$, GEMS $24.0<m_r<26.0$ and UVUDF $26.0<m_r<28.0$ data.}
\label{fig:magDistFaint}
\end{figure}

\begin{figure}
\centering
\includegraphics[width=0.45\textwidth]{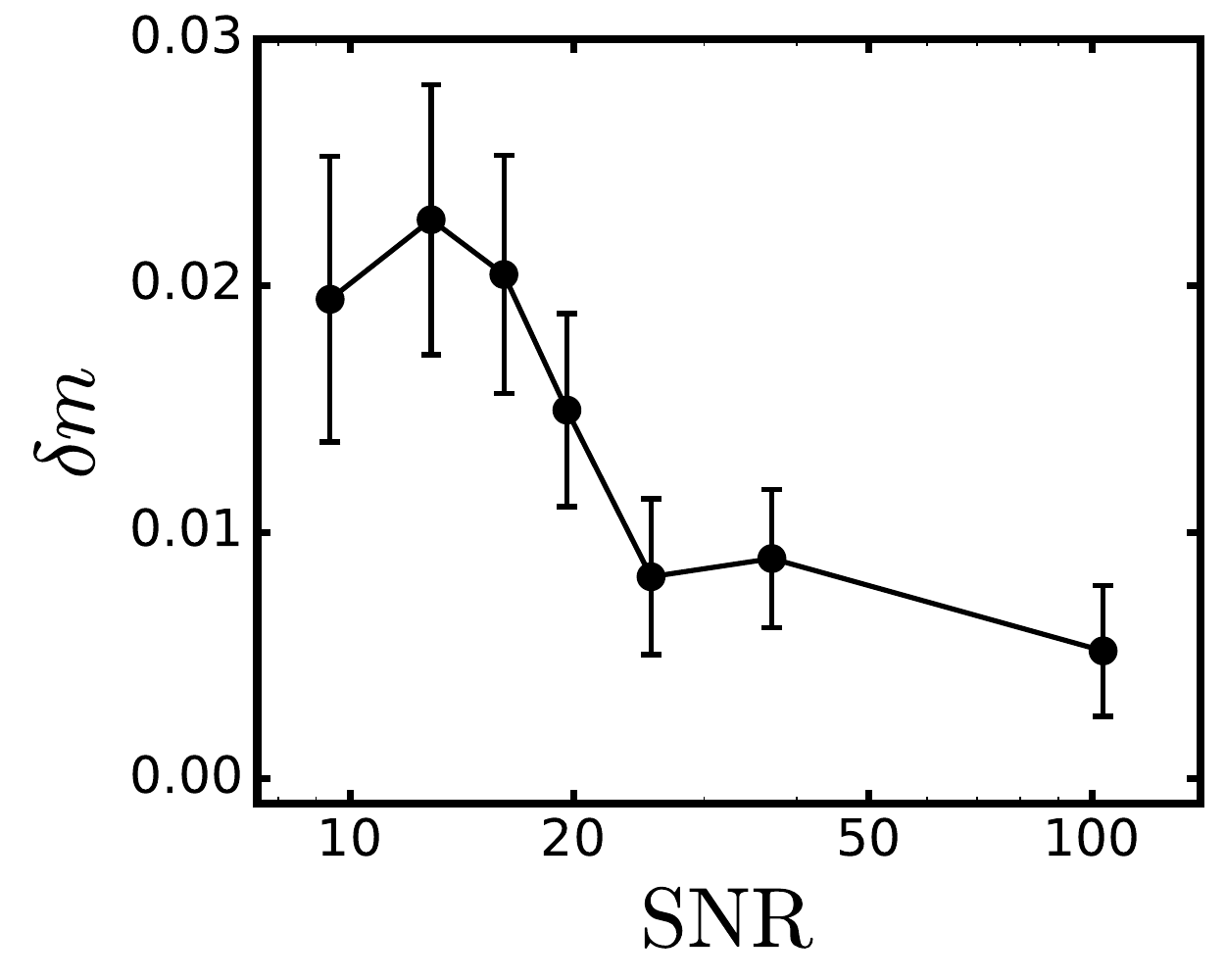}
\caption{Residual multiplicative bias $\delta{m}$ due to the presence of undetected galaxies 
as a function of SNR.}
\label{fig:deltaBiasSNR}
\end{figure}

For objects fainter than the limiting depth, \citet{2015MNRAS.449..685H} found that they 
make the multiplicative bias of brighter galaxies underestimated in the cluster environment 
because they are likely to be blenders or skew the background noise. 
\citet{2017MNRAS.468.3295H} further analyzed the issue, and found that the multiplicative 
bias is affected by both the size distribution and count slope of the undetected 
galaxies. \citet{2017MNRAS.467.1627F} also discussed the effect in KiDS simulation and 
found a negligible bias compared to the survey statistical uncertainties.

To mimic the realistic magnitude distribution of the undetected galaxies in VOICE observation, photometric measurements 
from other deeper imaging  are included, which are \textit{HST}/ACS $F606W$-band data from 
UVUDF \citep{2015AJ....150...31R}  and GEMS \citep{2004ApJS..152..163R,2012ApJS..200....9G}, 
Subaru/SuprimeCam $r$-band data from COSMOS2015 catalog \citep{2016ApJS..224...24L}. 
These filters are analogous to the OmegaCam $r$ filter. Figure \ref{fig:magDistFaint} shows their
number density distributions of galaxies as a function of magnitude $m$. It can be seen that they are consistent 
for magnitude brighter than 25.0\,mag. A second order polynomial is adopted 
to fit the distribution using VOICE counts between $20.0<m_r<24.0$, GEMS counts between $24.0<m_r<26.0$ 
and UVUDF counts between $26.0<m_r<28.0$. The least-square result is 
\begin{eqnarray}
\log(n) = -15.012 + 0.947m_r - 0.013m_r^2, \nonumber
\end{eqnarray}
where $n$ is the number of galaxies per square arcminute in a given magnitude bin with 
width of 0.2\,mag. In our simulation, we truncate the magnitude of undetected galaxies to 28.0\,mag, 
and restrict their bright-end to 25.0\,mag which is approximately equal to the maximal value
in the distribution of VOICE catalog, as depicted  in Figure \ref{fig:magDistFaint}. Consequently, 
the total number density of these undetected galaxies is as many as 185 per square arcminute.

Unlike the simulation of the detected galaxies as described in Section \ref{sec:ImgSim},   
the celestial positions of the undetected galaxies are randomly assigned. The size 
and intrinsic ellipticity are drawn from the same prior distributions as stated in 
\citet{2013MNRAS.429.2858M}. Since these galaxies are below the noise level even in 
the stacked image, extra shear components are not expected to contribute significant systematics.
Therefore, null shear is assigned to these galaxies. Furthermore, to save simulation time, 
we do not generate the undetected galaxy images chip by chip as for the procedure for the detectable ones. 
Instead, we sprinkle them to a noiseless image mosaic centered at the same celestial position as that of 
the CDFS field. The PSF model is assumed to be Gaussian with constant 
FWHM fixed to the median value of the observation. Finally, for a given CCD chip (or sky coverage), 
we extract the corresponding sub-image from this image mosaic, and then add it to the previously 
simulated image.

We follow the same steps as presented in previous sections to perform shear measurements and 
bias analyses for the detected galaxies using the new set of images. Compared to the results 
derived from no-faint-galaxies simulation, the multiplicative bias of the entire sample increases only 
by 0.003, while the additive bias shows negligible change. Figure \ref{fig:deltaBiasSNR} shows the 
residual multiplicative bias $\delta{m}$ as a function of SNR. Here $\delta{m}$ is defined as 
$[(\delta{m_1})^2+(\delta{m_2})^2]^{1/2}$, where $\delta{m_i}$ represents 
the difference of multiplicative biases between the two sets of simulation. Overall, our result indicates that 
galaxies with lower SNR (or fainter magnitude) suffer more from the 
undetected galaxies. Since the amplitude is well below the residual bias we achieve in 
Section \ref{sec:BiasCal}, we claim that the sensitivity of the multiplicative bias to the undetected 
galaxies for our simulation is insignificant. However, as illustrated in Figure \ref{fig:deltaBiasSNR} the impact 
of undetected galaxies has to be taken into account for more accurate shear measurements as required by 
future large and deep surveys, especially for galaxies with low SNRs.

\begin{figure}
\centering
\includegraphics[width=0.45\textwidth]{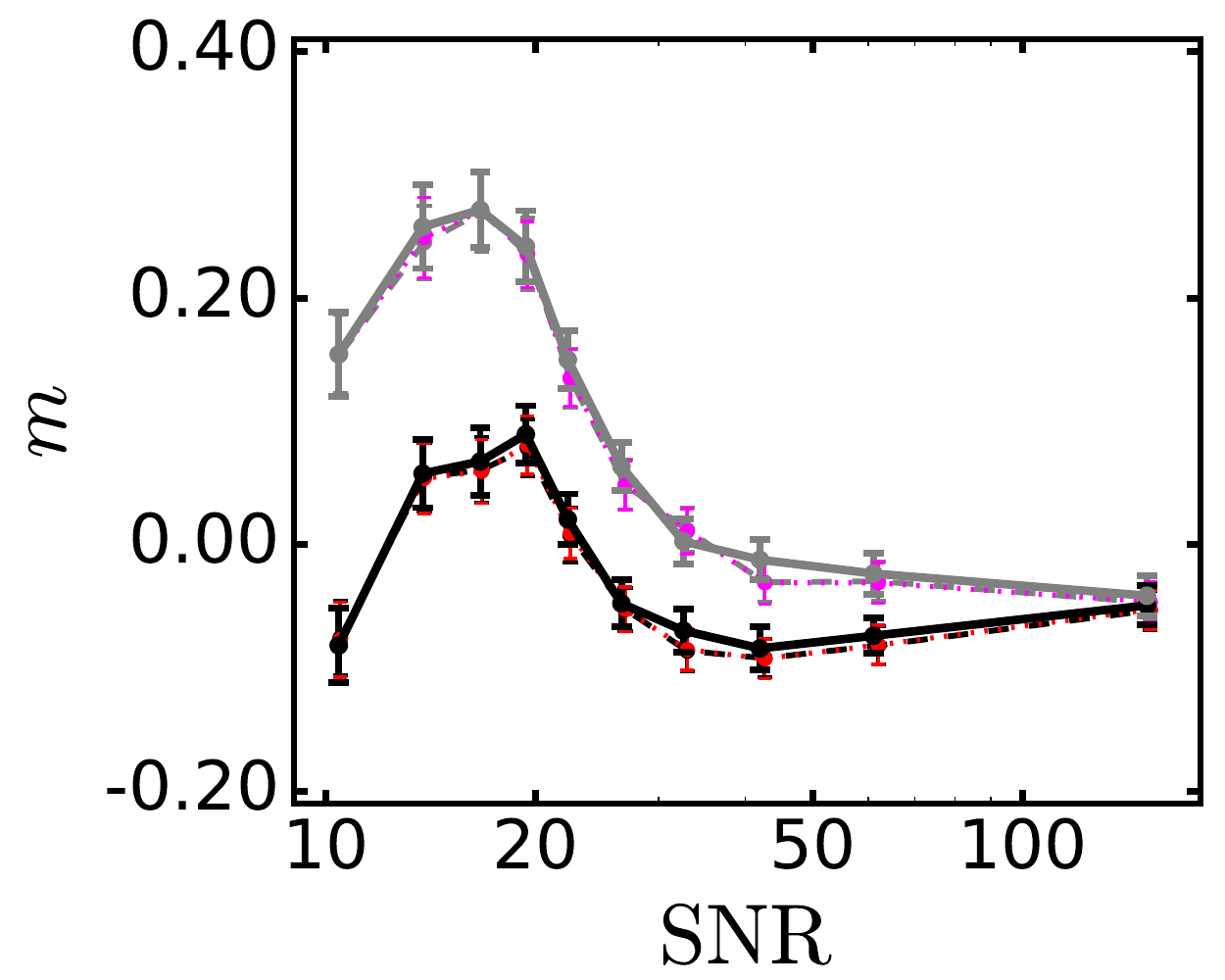}
\caption{The multiplicative bias as a function of SNR. The black and 
grey dashed lines represent the two bias components $m_1$ and 
$m_2$ derived from the entire simulation sample, while the solid 
lines are the corresponding components after rejecting the neighboring  
galaxies. The dotted lines indicate the average bias derived from the 
subsamples of galaxies by repeated sampling 100 times.}
\label{fig:SNRblend}
\end{figure}

\begin{table}
\centering
\caption{The ellipticity dispersions of the neighboring galaxies in the 
four fields. Column 1 represents the maximum distance between any galaxy pair.
The number of galaxies with non-zero weight and zero flag is shown in Column 2, while 
Column 3 indicates the fraction relative to the neighboring galaxies without any cut. 
Column 4\&5 are the dispersions of the two ellipticity components.}
\label{tab:ellDis}
\begin{tabular}{rrrrr} 
\hline
$r$ & $n_\mathrm{gal}$ & fracion & $\sigma_{e_1}$ & $\sigma_{e_2}$ \\
\hline
1.0\arcsec & 124     & 33.6\% & 0.403 & 0.421 \\
2.0\arcsec & 1858   & 7.8\% & 0.333 & 0.350 \\
3.0\arcsec & 25746 & 31.6\% & 0.307 & 0.309 \\
\hline
\end{tabular}
\end{table}

\begin{figure}
\centering
\includegraphics[width=0.45\textwidth]{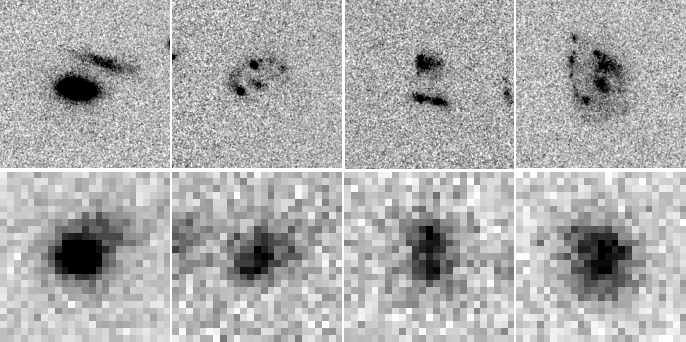}
\caption{Examples of four very close neighbors in CDFS1/GEMS field. The top panel 
shows galaxies observed in GEMS survey, while the corresponding stacked images 
from the VOICE survey are displayed in bottom panel. The size of each stamp is 
$5\arcsec\times5\arcsec$, centered on the target galaxy in VOICE image.}
\label{fig:blend}
\end{figure}

\subsection{Impact of Blending Galaxies}
As discussed above, the simulation strategy in this work enables 
us to study the effect of the neighboring galaxies on the measured shear. These galaxies 
can be either physically related neighbors with similar shear or projected close 
pairs but with different redshifts and shape distortions. Although \texttt{LensFit} has encoded 
an algorithm to deal with them \citep{2013MNRAS.429.2858M}, potential bias is still inevitable 
in the measured shear due to the inappropriate modeling of the surface brightness 
distributions in the overlapped regions. In this section, we mainly concentrate on their 
contribution to the multiplicative bias.

First of all, we compare the ellipticity dispersions of the real galaxies in the parent sample
which is constructed from observation and used as simulation input (see Sec \ref{sec:mcat}).
As shown in Table~\ref{tab:ellDis}, the dispersion increases as the neighbors get closer to each 
other, and is always larger than that of all galaxies in the parent sample. We note that a large 
fraction of the neighbors with separation less than $1.0\arcsec$ have non-zero weight. Careful analysis shows that \texttt{LensFit} 
regards most of these close galaxy pairs as single and extended galaxies. The resulted ellipticity dispersion  
is almost 38.0\% larger than that of the parent sample. Clearly, these galaxies should be excluded from further shear 
analyses. For galaxy pairs with the separation less than $r=3.0\arcsec$, about 31.6\% of them have shape measurements 
with non-zero weight. Their number density is about 1.3 per square arcmin, and the ellipticity dispersion is about 3.4\% larger 
than that of the full shear sample. Rejecting these galaxies results in a $\sim$8.2\% decrease of the effective number density of 
galaxies. On the other hand, because of relatively large statistical uncertainties of VOICE data,  the results of cosmic 
shear two-point correlations have  negligible changes if these galaxies are excluded.  

We further quantify their impact on the multiplicative bias using the simulated shear catalogs. Considering different sizes of galaxies, 
to reflect better the blending effect, we redefine galaxy neighbors using an adaptive scheme with a separation less than four times 
of the sum of their major axis scalelengths.  Under this definition, the fraction of neighbors with shape measurements 
is about 6\%. A clean sample is then constructed by rejecting these 
neighbors from the full simulation sample. The solid lines in Figure~\ref{fig:SNRblend} show the two multiplicative bias components, 
derived from the clean sample, as a function of SNR. For comparison, the bias components from 
the full sample are illustrated as dashed lines. It is seen that the multiplicative biases of the two components 
become systematically smaller for the clean sample than those of the full sample, although the differences 
are not very considerable. It is interesting to note that the differences are larger for galaxies with high SNR. 
This is because the SNRs of the neighboring galaxies are systematically overestimated, and their effect on 
the bias is therefore more significant at high SNRs. We find that their median SNR is 33.0, comparing to 24.2 of 
the full sample.  

The clean sample contains slightly less galaxies. To see if the number change can affect the bias calibration, 
we randomly select an equal number of galaxies as the clean sample from the full simulation 
sample, and estimate the multiplicative bias. The sampling procedure is repeated 
by one hundred times. The dotted lines in Figure~\ref{fig:SNRblend} show the average of the bias 
as a function of SNR. We see that the results are basically the same as those of the full sample, 
showing that the blending effect does contribute to the differences between the clean and the full sample. 
The differences are at the level of $0.002$. 

Besides the neighbors that can be unambiguously identified, there are pairs that are so close and misidentified 
as single objects. This is particularly the case for ground-base observations. To check this for the VOICE sample, 
we use the data of Galaxy Evolution From Morphology And SEDs (GEMS; \citet{2004ApJS..152..163R}) observed 
using Hubble space telescope. The overlapped area between VOICE and GEMS is about 800\,arcmin$^2$. 
We identified 2185 such blenders down to magnitude of 26.0\,mag 
in our parent sample by following the similar method in \citet{2016ApJ...816...11D}. This accounts for about 5.0\% of 
the total number of galaxies in the overlapped area. We find that 68\% of them 
have shape measurements with dispersion of 0.33 for the two ellipticity components. Their weighted 
number density is 0.92 per arcmin$^2$. Figure~\ref{fig:blend} exhibits four typical examples 
of these blenders. Apparently, they are observed as multiple objects in the GEMS survey, and  
show diverse morphologies. However, in the VOICE observations, they are identified as single galaxies. 
Their shear measurements using VOICE data cannot be correct, and thus should be excluded. 
We do not expect that they affect significantly our VOICE shear analyses because of the large 
statistical errors. For the upcoming deep and wide ground-based surveys, however, we do need to consider 
such blenders, and quantify carefully how they affect the weak lensing cosmological studies. 

In summary, galaxies fainter than the detection limit and the blending effect from neighboring galaxies 
contribute to the multiplicative bias at the level of less than $\sim$0.5\%. The small differences between 
our simulation catalog and the observed data do not induce noticeable biases (less than $1\%$) considering 
the statistical uncertainties of the VOICE shear sample. Our final multiplicative shear calibration residual 
is $\sim$3\%, which reflects mainly the statistical errors. 

\section{Summary}\label{sec:Con}
The VOICE survey has observed
$\sim$ 4\, deg$^2$ in the CDFS field in $ugri$ optical bands using VST/OmegaCam. 
After a cut in FWHM $<$ 0.9 arcsec, the survey consists of more than a hundred 
exposures for each tile, and the depth is about 1.2 magnitude deeper than that of KiDS 
survey. We have performed shear measurements, and obtained 
an effective number density of galaxies $n_g\sim 16.35 \hbox{ arcmin}^{-2}$.  
In the work, we perform detailed shear bias calibrations for the VOICE survey based on $r$-band image 
simulations. Many observational conditions, such as the 
dithering pattern, background noise, celestial positions and brightness of 
the detected objects, have been taken into account in the simulations in order to mimic the real 
observations. The \texttt{PSFEx} package is used to model the spatially varied PSF 
in every exposure. The simulated single exposure images are 
generated by the \texttt{Galsim} toolkit, and the galaxy shapes are measured 
by \texttt{LensFit}, a Bayesian fitting code that has been extensively applied 
to many other large surveys, such as CFHTLenS, KiDS and RCSLenS. 
Overall, our simulations present good agreements with the characteristics of observations, 
especially the distributions of the PSF parameters. We notice that some small and faint galaxies are 
missing in our simulations comparing to the real observations. We argue that they should not 
affect our shear calibration significantly given the relatively low total number of galaxies in the VOICE survey.  
We apply the bin-matching method to the 
SNR and size surface to calibrate the bias of the simulation data. The final residual 
multiplicative bias can reach to an accuracy of 0.03 with negligible additive bias in 
different SNR and size bins. The average residual bias of the full sample is consistent with zero.

Our studies demonstrate the applicability of  \texttt{Lensfit} for shear measurements  to data with 
more than a hundred exposures. The image simulation analyses show that the change of the 
deblending threshold from the fiducial $2\sigma$ to $5\sigma$ does not introduce considerable issues. 

We further discuss the sensitivity of the bias calibration to the undetected and blending  
objects. The undetected objects are likely to skew the background noise so that they 
can potentially bias the shape measurements of galaxies, especially those with low 
SNR. Taking the depth and noise level, and the relatively large statistical uncertainties into account, 
we find that the impact of the undetected galaxies is negligible for the VOICE survey. Additionally,  we highlight the 
bias resulting from galaxy blending effect. Although a large fraction of neighboring galaxies 
has been excluded by \texttt{LensFit}, there are still 31.6\% of neighboring galaxies with separation less 
than $3\arcsec$ having shape measurements. The ellipticity dispersion of them is 3.4\% larger than 
the average value of the parent sample, and the weighted number density is as large as 1.3 per 
square arcmin. Considering galaxy pairs with an adaptive separation less that four times of the sum 
of their major axes, we find that $\sim$6\% of them have shear measurements, and they contribute 
to additional $\sim$0.2\% multiplicative bias.  With the increase of depth and sensitivity, many weak lensing surveys, 
such as HSC and LSST, aiming to achieve much more accurate weak lensing studies than that of VOICE, 
have to deal with the blending effect more carefully.

\section*{acknowledgements}
We thank Jun Zhang for helpful comments on the image simulation.  
The research is supported in part by NSFC of 
China under the grants 11333001, 11173001 and 11653001. 
L.P.F. acknowledges the
support from NSFC grant 11673018, 11722326 \& 11333001, STCSM grant 16ZR1424800 
and SHNU grant DYL201603.
X.K.L. acknowledges the support from YNU Grant KC1710708 and General Financial Grant from China 
Postdoctoral Science Foundation with Grant No. 2016M591006.
M.R. acknowledges the PRIN MIUR ``Cosmology and Fundamental Physics: illuminating the Dark Universe with Euclid".
M.V. acknowledges support from the European Commission Research Executive Agency
(FP7-SPACE-2013-1 GA 607254), the South African Department of Science and Technology
(DST/CON 0134/2014) and the Italian Ministry for Foreign Affairs and International
Cooperation (PGR GA ZA14GR02).
M.P. acknowledges support from ASI-INAF grant 2017-14-H.O: Studies for the High Energy Community:
Data Analysis, Theory and Simulations. 
Support for G.P. is provided by the Ministry of Economy, Development, and Tourism's Millennium Science 
Initiative through grant IC120009, awarded to The Millennium Institute of Astrophysics, MAS

Based on data products from observations made with ESO Telescopes at the  
Paranal Observatory under ESO programme ID 179.A-2005 and on
data products produced by TERAPIX and the Cambridge Astronomy Survey
Unit on behalf of the UltraVISTA consortium.

Some/all of the data presented in this paper were obtained from the Mikulski Archive 
for Space Telescopes (MAST). STScI is operated by the Association of Universities for 
Research in Astronomy, Inc., under NASA contract NAS5-26555. Support for MAST for 
non-HST data is provided by the NASA Office of Space Science via grant NNX09AF08G 
and by other grants and contracts.


\end{document}